\documentclass{aastex63}

\newcommand{\xie}{\color{black}}
\newcommand{\chen}{\color{black}}
\newcommand{\chendichang}{\color{black}}

\newcommand{\respondtoxiang}{\color{black}}

\usepackage{amsmath}
\usepackage{amssymb}
\usepackage{color}
\usepackage{url}
\usepackage{ulem}
\usepackage{multirow}
\usepackage{graphics}
\usepackage{longtable}
\usepackage{graphicx}

\begin{document}

\title{Planets Across Space and Time (PAST). \uppercase\expandafter{\romannumeral2}: Catalog and Analyses of the  LAMOST-Gaia-Kepler Stellar Kinematic Properties}

\correspondingauthor{Ji-Wei Xie}
\email{jwxie@nju.edu.cn}


\author{Di-Chang Chen}
\affiliation{School of Astronomy and Space Science, Nanjing University, Nanjing 210023, China}
\affiliation{Key Laboratory of Modern Astronomy and Astrophysics, Ministry of Education, Nanjing 210023, China}

\author{Jia-Yi Yang}
\affiliation{School of Astronomy and Space Science, Nanjing University, Nanjing 210023, China}
\affiliation{Key Laboratory of Modern Astronomy and Astrophysics, Ministry of Education, Nanjing 210023, China}

\author{Ji-Wei Xie}
\affiliation{School of Astronomy and Space Science, Nanjing University, Nanjing 210023, China}
\affiliation{Key Laboratory of Modern Astronomy and Astrophysics, Ministry of Education, Nanjing 210023, China}

\author{Ji-Lin Zhou}
\affiliation{School of Astronomy and Space Science, Nanjing University, Nanjing 210023, China}
\affiliation{Key Laboratory of Modern Astronomy and Astrophysics, Ministry of Education, Nanjing 210023, China}

\author{Subo Dong}
\affiliation{Kavli Institute for Astronomy and Astrophysics, Peking University, Beijing 100871, China}

\author{Zheng Zheng}
\affiliation{Department of Physics and Astronomy, University of Utah, Salt Lake City, UT 84112}

\author{Jing-Hua Zhang}
\affiliation{National Astronomical Observatories, Chinese Academy of Sciences, Beijing 100012, China}

\author{Chao Liu}
\affiliation{Key Lab of Space Astronomy and Technology, National Astronomical Observatories, CAS, 100101, China}
\affiliation{University of Chinese Academy of Sciences, Beijing, 100049, China}

\author{Hai-Feng Wang}
\affiliation{South-Western Institute for Astronomy Research, Yunnan University, Kunming, 650500, China; LAMOST Fellow}

\author{Mao-Sheng Xiang}
\affiliation{National Astronomical Observatories, Chinese Academy of Sciences, Beijing 100012, China}
\affiliation{Max-Planck Institute for Astronomy, Königstuhl 17, D-69117 Heidelberg, Germany}

\author{Weikai Zong}
\affiliation{Department of Astronomy, Beijing Normal University, Beijing 100875, China}

\author{Yang Huang}
\affiliation{South-Western Institute for Astronomy Research, Yunnan University, Kunming, 650500, China}

\author{Ali Luo}
\affiliation{National Astronomical Observatories, Chinese Academy of Sciences, Beijing 100012, China}



\begin{abstract}
The Kepler telescope has discovered over 4,000 planets (candidates) by searching $\sim$ 200,000 stars over a wide range of distance (order of kpc) in our Galaxy. Characterizing the kinematic properties (e.g., Galactic component membership and kinematic age) of these {\chen Kepler targets} (including the {\chendichang planet candidate host}s) is the first step towards studying Kepler planets in the Galactic context, which will reveal fresh insights into planet formation and evolution. 
In this paper, the second part of the Planets Across the Space and Time (PAST) series, by {\chen combining} the data from LAMOST and Gaia and then applying the revised kinematic methods from PAST \uppercase\expandafter{\romannumeral1}, we present a catalog of kinematic properties (i.e., Galactic positions, velocities, and the relative membership probabilities among the thin disk, thick disk, Hercules stream, and the halo) as well as other basic stellar parameters for {\chen 35,835} Kepler stars.
Further analyses of the LAMOST-Gaia-Kepler catalog demonstrate that our derived kinematic age reveals the expected stellar activity-age trend.  
Furthermore, we find that the fraction of thin (thick) disk stars increases (decreases) with the transiting planet multiplicity ($N_{\rm p}=$ 0, 1, 2 and $3+$) and the kinematic age decreases with $N_{\rm p}$, which could be a consequence of the dynamical evolution of planetary architecture with age. 
The LAMOST-Gaia-Kepler catalog will be useful for future studies on the correlations between the exoplanet distributions and the stellar Galactic environments as well as ages.

\end{abstract}

\section{introduction}
With the discovery of thousands of exoplanets, the scope of planetary research has been expanding from the Solar system/neighborhood to a wider range of our Milky Way Galaxy. Understanding how planetary properties depend on the Galactic environment and age will provide crucial insights on planet formation and evolution.
Towards a Galactic census of planets, we have started a research project, dubbed Planets Across Space and Time (PAST).
In the Paper \uppercase\expandafter{\romannumeral1} \citep[here after PAST \uppercase\expandafter{\romannumeral1},][]{2021ApJ...909..115C} 
of the PAST series,  we revisited the kinematic methods for classification of Galactic components and extended the applicable range of the methods from $\sim 100-200$ pc to  $\sim1.5$ kpc to cover most of {\chendichang planet candidate host} stars. 
{\chen Additional}, we revised the Age-Velocity dispersion Relation (AVR), which allows us to derive kinematic age with a typical uncertainty of $\sim 10-20\%$ for an ensemble of stars.
Here, in the second paper of PAST (PAST \uppercase\expandafter{\romannumeral2}), we apply the revised kinematic methods and AVR to a large homogeneous sample of stars with the observational synergy among LAMOST \citep[the Large Sky Area Multi-Object Fiber Spectroscopic Telescope, also known as Goushoujing Telescope; spectroscopy,][]{1996ApOpt..35.5155W,2004ChJAA...4....1S,2012RAA....12.1197C,2012RAA....12..723Z,2012RAA....12.1243L,2018ApJS..238...30Z}, Gaia \citep[astrometry,][]{2018A&A...616A...1G,2018A&A...616A..11G} and {\chen Kepler \citep[photometry,][]{2017ApJS..229...30M}}.  

The Kepler mission has provided an unprecedented legacy sample for stellar astrophysics and exoplanet science, thanks to the long-term baseline, high-precision observations of a large amount ($\sim$ 200,000) of stars  \citep{2011AJ....142..112B,2014ApJS..211....2H,2017ApJS..229...30M}, which discovered 2,348 confirmed planets and 2,368 candidates \citep[data from NASA Exoplanet Archive, EA hereafter;][] {2013PASP..125..989A}, contributing the majority of confirmed exoplanets and candidates.
One of advantages of the Kepler sample is that it is suitable for studies on planetary systems in the Galactic context because they spread over different Galactic environments in a wide region up to several kpc in our Milky way \citep[Fig 13 of PAST \uppercase\expandafter{\romannumeral1},][]{2021ApJ...909..115C}.
In order to have a reliable Galactic census of planets, one needs accurate characterizations of the Kepler stars (not just the {\chendichang planet candidate host}s). 
For most Kepler stars, some stellar parameters can be measured in relatively high precision, e.g., 7\% for mass, 4\% for radius and 112 K for effective temperature, while the uncertainty of stellar age is relatively large, e.g., 56\% on average from isochrone fitting \citep{2020AJ....159..280B}.  
The kinematic properties (e.g., Galactic velocities and thin/thick disk memberships) of Kepler stars, however, still remain to be characterized.
Furthermore, as we will show below, kinematic ages derived from Galactic velocity dispersions with AVR can be an important complementary to ages derived from other methods.
Therefore, in this paper, we focus on the kinematic characterization of the Kepler stars. 



To obtain the Galactic component memberships and kinematic age, we only need the 3D space motions without involving stellar evolutionary mode \citep[e.g.][]{2003A&A...410..527B,2010ARA&A..48..581S,2014A&A...565A..89B}, which can be derived from radial velocity plus five astrometric parameters, i.e., parallax, celestial coordinates (Right Ascension and Declination) and the corresponding proper motions. 
Fortunately, the second Data release (DR2) of  Gaia \citep[e.g.,][] {2016A&A...595A...1G,2018A&A...616A...1G,2018A&A...616A..11G} has provided the five astrometric parameters for more than 1.3 billion stars, including $\sim 180,000$ Kepler stars \citep{2020AJ....159..280B}.
However, Gaia DR2 provides radial velocities only for bright stars (about 7.2 million with a magnitude of $G \sim 4-13$) and thus only include $\sim 10\%$ of Kepler stars.
Thus to make up for that, we rely on the spectroscopic data from LAMOST,
which provides radial velocities for over 30\% of all Kepler targets with no bias toward Kepler {\chendichang planet candidate host}s \citep{2015ApJS..220...19D,2016ApJS..225...28R,Fu.2020RAA....20..167F}. 
Besides, LAMOST also provides other physics parameters (e.g. $\rm [Fe/H]$, $\rm [\alpha/Fe]$, and the magnetic S index), which will enrich our knowledge of the stars and benefit the studies concerning stellar physical properties.

In this paper, we kinematically characterize {\chen 35,835} Kepler stars based on their astrometry and radial velocities provided by Gaia and LAMOST. 
Specifically, 
in section \ref{sec.samp}, we describe how to collect the stellar sample from Kepler, LAMOST and Gaia data.  
In section \ref{sec.meth}, we briefly introduce the kinematic methods to identify {\respondtoxiang Galactic} components (e.g., thin/thick disk) and to derive kinematic ages using the AVR.
In section \ref{sec.res.cat}, we conduct a catalog of kinematic properties for {\chen 35,835} LAMOST-Gaia-Kepler stars.
In section \ref{sec.cat.analy}, we carry out some relative explorations based on our catalog.
Finally, we summarize in section \ref{sec.summary}.

\section{Data Collections}
\label{sec.samp}
This section describes how we constructed the stellar sample from Gaia, Kepler and LAMOST for kinematic characterizing. 

\subsection{Kepler: Exoplanet Transit Surveys}
We initialized our sample from the Kepler Data Release 25 (DR25) catalog, which contains 197,096 Kepler target stars as well as 8,054 Kepler Objects of Interest (KOIs) \citep{2017ApJS..229...30M}. Here, we excluded KOIs  flagged  by  ‘False Positive  (FAP)’,  leaving 4,034 planets (candidates).
We also removed potential binaries by eliminating stars with Gaia DR2 re-normalized unit-weight error (RUWE) $>1.2$ \citep{2018AJ....156..195R,2020AJ....160..108B} as additional motions caused by binary orbits could affect the results of kinematic characterization.
After these cuts, we are left with 175,280 Kepler stars and 3,620 planets (candidates).


\subsection{Obtaining Five Astrometric Parameters from Gaia}
\label{sec.samp.5ast}
To obtain the astrometric parameters for Kepler stars, we cross-matched them with Gaia DR2, which includes  positions on the sky $(\alpha, \delta)$, parallaxes, and proper motions ($\mu_{\alpha}, \mu_{\delta}$) for more than 1.3 billion stars with a limiting magnitude of G $ = 21$ and a bright limit of G $\approx 3$ \citep{2018A&A...616A...1G}.
The cross-matching was done by adopting the X match service of the Centre de Donnees astronomiques de Strasbourg (CDS, http://cdsxmatch.u-strasbg.fr, \cite{2012ASPC..461..291B}). 
After inspecting the distribution of separations, we select the separation limit of the cross-matching as where the distribution of separations displayed a minimum,  $\sim1.5$ arcseconds.
To ensure that the matched stars are of similar brightness, we also make a magnitude cut.
The magnitude limit was set by inspecting the distribution of magnitude difference, which is 2 mag in Gaia G mag.
For multiple matches satisfied these two criteria for the same star, we kept the one with the smallest angular separation.
After these selections, we obtained 163,454 stars and 3,409 candidates.

\subsection{Obtaining RV from LAMOST}
\label{sec.samp.rv}
Next, to obtained RV data, we rely on the spectroscopic data from LAMOST \citep{2012RAA....12.1197C,2012RAA....12..723Z}.
The LAMOST DR4 value-added catalog contains parameters derived from a total of 6.5 million stellar spectra for 4.4 million unique stars  \citep{2017MNRAS.467.1890X}. 
RVs, $T_{\rm eff}$, $\log g$, and $\rm [Fe/H]$ have been deduced using both the official LAMOST Stellar parameter Pipeline \citep[LASP; ][]{2011RAA....11..924W} and the LAMOST Stellar Parameter Pipeline at Peking University \citep[LSP3; ][]{2015MNRAS.448..822X}. 
The typical uncertainties for RVs, $T_{\rm eff}$, $\log g$, and $\rm [Fe/H]$ are 5.0 $\rm km \ s^{-1}$, 150 K, 0.25 dex, and 0.15 dex respectively. 

We cross-matched the foregoing sample with LAMOST DR4 value-added catalog by using CDS with the same procedure detailed in section \ref{sec.samp.5ast}. By inspecting the distribution of separations and magnitude difference, the separation limit and magnitude cut were set as 5 arcseconds and 2.3 mag. 
Besides, we applied a quality cut of $\rm SNR>10$.
To ensure the reliability of RV, we also cross-matched with the LAMOST DR7 catalog {\chen (http://dr7.lamost.org/)} and remove the stars when the differences in RVs are larger than three times of the uncertainties.
We only kept stars with a distance less than 1.54 kpc to the sun, which is the applicable limit of the revised kinematic characteristics and AVR in PAST \uppercase\expandafter{\romannumeral1} \citep{2021ApJ...909..115C}.
Finally we obtained a LAMOST-Gaia-Kepler sample of {\chen 35,835} stars and {\chen 1,060} planets.
In Table \ref{tab:planetnumberprocedure}, we summarize the composition of the sample after each step mentioned above.

\begin{table}[!t]
\centering
\caption{Construction of the LAMOST-Gaia-Kepler star sample.}
{\footnotesize
\label{tab:planetnumberprocedure}
\linespread{1.8}
\begin{tabular}{l|cc} \hline
Selection criteria & $N_{\rm s}$ & $N_{\rm p}$ \\\hline
Kepler DR25 & 197,096 & 8,054 \\
No FAP & ... & 4,034 \\
No binary & 175,280 & 3,620 \\
Cross-matched with Gaia & 163,454 & 3,409 \\
Cross-match with LAMOST DR4&  {\chen 55,020} & {\chen 1,202} \\
$\rm SNR>10$ & {\chen 54,337} & {\chen 1,183} \\ 
RV reliability cut & {\chen 49,261} & {\chen 1,098} \\ 
Distance $<1.54$ kpc & {\chen 35,835}  & {\chen 1,060} \\ \hline
\end{tabular}}
\flushleft
{\scriptsize
 $N_{\rm s}$ and $N_{\rm p}$ are the numbers of stars and planets during the process of sample selection in section \ref{sec.samp}.}
\end{table}


\section{Methods: Classification of Galactic Components and Estimation of Kinematic Ages} 
\label{sec.meth}
In this section, we describe how we distinguish star into different Galactic components (section \ref{sec.meth.class}) and estimate stellar ages (section \ref{sec.kine.age.err}) with revisiting velocity ellipsoid and AVR from PAST \uppercase\expandafter{\romannumeral1} based on data from LAMOST and Gaia DR2.

\subsection{Space Velocities and {\respondtoxiang Galactic} Orbits}
\label{sec.meth.space}
We calculated the 3D Galactocentric cylindrical coordinates $(R, \theta, Z)$ by adopting a location of the Sun of $R_\odot$= 8.34 kpc \citep{2014ApJ...783..130R} and $Z_\odot$ = 27 pc \citep{2001ApJ...553..184C}. 
By adopting the formulae and matrix equations presented in  \cite{1987AJ.....93..864J}, we calculate the {\respondtoxiang Galactic} rectangular velocities relative to the Sun $(U, V, W)$ and their errors for the LASMOT-Gaia-Kepler sample.
Here $U$ is positive when pointing to the direction of the {\respondtoxiang Galactic} center, $V$ is positive along the direction of the Sun {\xie orbiting around the {\respondtoxiang Galactic} center}, and $W$ is positive when pointing towards the North {\respondtoxiang Galactic} Pole. 
To obtain the  Galactic rectangular velocities relative to the local standard of rest (LSR) $(U_{\rm LSR}, V_{\rm LSR}, W_{\rm LSR})$, we adopted the solar peculiar motion [$U_\odot$, $V_\odot$, $W_\odot$] = [9.58, 10.52, 7.01] $\rm km \ s^{-1}$  \citep{2015ApJ...809..145T}. 
Cylindrical velocities $V_R$, $V_\theta$, and $V_Z$ are defined as positive with increasing $R$, $\theta$, and $Z$, with the latter towards the North {\respondtoxiang Galactic} Pole.

\subsection{Classification of Galactic Components}
\label{sec.meth.class}
In this section, by adopting the wide-used kinematic approaches \citep{2003A&A...410..527B,2014A&A...562A..71B}, we classify stars into different Galactic components (i.e.,  thin disk,  thick disk, halo and Hercules stream) based on their Galactic positions and velocities.
The kinematic methods assumes the Galactic velocities to the LSR ($U_{\rm LSR}, V_{\rm LSR}, W_{\rm LSR}$) have multi-dimensional Gaussian distributions:
\begin{equation}
\begin{aligned}
&f (U,V,W) = k \times {\rm exp} \\
&\left(-\frac{(U_{\rm LSR}-U_{asym})^2}{2{\sigma_U}^2}
-\frac{(V_{\rm LSR}-V_{asym})^2}{2{\sigma_V}^2}
-\frac{W_{\rm LSR}^2}{2{\sigma_W}^2}\right),
\label{fUVW}
\end{aligned}
\end{equation}
where $\sigma_U$, $\sigma_V$, and $\sigma_W$ are the characteristic velocity dispersions, and $V_{\rm asym}$ and $U_{\rm asym}$ are the asymmetric drifts for different components (the thin disk, the thick disk, the halo, and the Hercules stream). The normalization coefficient are defined as
\begin{equation}
k = \frac{1}{(2\pi)^{3/2}\sigma_U\sigma_V\sigma_W}.
\label{kUVW}
\end{equation}

The relative probabilities  between two different components, i.e.,  the thick-disk-to-thin-disk $(TD/D)$, thick-disk to halo $(TD/H)$, the Hercules-to-thin-disk $(Herc/D)$, and the Hercules-to-thick-disk $(Herc/TD)$ can be calculated as 
\begin{alignat}{2}
 \frac{TD}{D} &=\frac{X_{\rm TD}}{X_{\rm D}} \cdot \frac{f_{\rm TD}}{f_{\rm D}},  &\quad \frac{TD}{H} &=\frac{X_{\rm TD}}{X_{\rm H}} \cdot \frac{f_{\rm TD}}{f_{\rm H}}, \\ 
 \frac{Herc}{D} &=\frac{X_{\rm Herc}}{X_{\rm D}} \cdot \frac{f_{\rm Herc}}{f_{\rm D}},  &\quad \frac{Herc}{TD} &=\frac{X_{\rm Herc}}{X_{\rm TD}} \cdot \frac{f_{\rm Herc}}{f_{\rm TD}},
\end{alignat}
\label{eqTDD}
where X is the fraction of stars for a given component.

Here we adopt {\chen the revised kinematic characteristics} and X factor of different Galactic components in PAST \uppercase\expandafter{\romannumeral1} \citep{2021ApJ...909..115C} and calculate the above Galactic component membership probabilities for the LAMOST-Gaia-Kepler stars.
Then we classified them into different Galactic components by adopting the same criteria as in \cite{2014A&A...562A..71B}, which are: \\
(1) thin disk: $TD/D<0.5 \,\,\&\,\, Herc/D<0.5$;\\
(2) thick disk: $TD/D>2 \,\,\&\,\, TD/H>1 \,\,\&\,\, Herc/TD<0.5$; \\
(3) halo: $TD/D>2 \,\,\&\,\, TD/H<1 \,\,\&\,\, Herc/TD<0.5$; \\ 
(4) Hercules: $Herc/D>1 \,\,\&\,\, Herc/TD>1$.

\begin{table}[!t]
\centering
\renewcommand\arraystretch{1.25}
\caption{Fitting parameters of the Age-Velocity dispersion relation from PAST \uppercase\expandafter{\romannumeral1} \citep{2021ApJ...909..115C}.}
{\footnotesize
\label{tab:AVRkb}
\begin{tabular}{l|cccccc} \hline
                & \multicolumn{2}{c}{---------~~$k \ \rm (km \ s^{-1})$~~---------} & \multicolumn{2}{c}{---------~~$\beta$~~---------}    \\ 
               &  value & 1 $\sigma$ interval &  value & 1 $\sigma$ interval  \\ \hline
    $U$  & $23.66$ & (23.07, 24.32) & $0.34$ & $(0.33,0.36)$  \\
    $V$  & $12.49$ & (12.05, 12.98) & $0.43$ & $(0.41,0.45)$  \\
    $W$  & $8.50$ & (8.09, 8.97) & $0.54$ & $(0.52,0.56)$ \\
    $V_{\rm tot}$ & $27.55$ & (26.84, 28.37) & $0.40$ & $(0.38, 0.42)$ \\ \hline
 
\end{tabular}}
\end{table}   

\subsection{Calculating Kinematic Age and Uncertainty}
\label{sec.kine.age.err}
As described in section 3.6 of PAST \uppercase\expandafter{\romannumeral1}, for a group of stars, the typical kinematic age can be derived by using the Age-velocity dispersion relation (AVR), which gives
\begin{equation}
{\rm Age_{kin}} \equiv t = \left(\frac{\sigma}{k\rm\, km \ s^{-1}}\right)^{\frac{1}{\beta}}\, \rm Gyr,
\label{eqkineage}
\end{equation}
where $\sigma$ is the velocity dispersion, which is defined as the root mean square of stellar Galactic velocity.
$k$, $\beta$ are the fitting coefficients of AVR.
By means of error propagation, the {\xie relative} uncertainty of kinematic age can be estimated as:
\begin{equation}
\begin{aligned}
\frac{\Delta t}{t}
        &= \sqrt{{(\frac{\partial \ln{t}}{\partial \beta} \Delta \beta)}^2
        +{(\frac{\partial \ln{t}}{\partial k} \Delta k)}^2+
        {(\frac{\partial \ln{t}}{\partial \sigma} \Delta \sigma)}^2}\\
      & = \sqrt{{\left(\ln \frac{t}{\rm Gyr}\right)^2 \left(\frac{\Delta \beta}{\beta}\right)^2
        +\frac{1}{\beta^2} \left(\frac{\Delta k}{k}\right)^2
        +\frac{1}{\beta^2} \left(\frac{\Delta\sigma}{\sigma}\right)^2}},
\label{eqkineageerr}
\end{aligned}
\end{equation}
where $\Delta$ represents the absolute uncertainty. 
Here we adopt the two coefficients ($k$, $\beta$) derived in PAST \uppercase\expandafter{\romannumeral1} \citep{2021ApJ...909..115C}, which are listed in Table \ref{tab:AVRkb}.

\section{ A Catalog of LAMOST-Gaia-Kepler stars with Kinematic Characterizations}
\label{sec.res.cat}

%

Applying the methods described in section \ref{sec.meth} to the LAMOST-Gaia-Kepler sample (section \ref{sec.samp}), we characterize the kinematic properties, e.g., Galactic orbit, velocities and the relative membership probabilities between different Galactic components ($TD/D$, $TD/H$, $Her/D$ and $Her/TD$) for the {\chen 35,835} LAMOST-Gaia-Kepler stars (Table \ref{tab:starcatalog}).
Among these stars, there are {\chen 764} stars hosting {\chen 1,060} planets.

Based on the derived relative membership probabilities between different Galactic components ($TD/D$, $TD/H$, $Her/D$ and $Her/TD$ in Table \ref{tab:starcatalog}), following the criteria as mentioned in section \ref{sec.meth.class}, we then classify the {\chen 35,835} stars into four Galactic components, i.e., thin disk, thick disk, Hercules stream and halo.  
For these stars not belonging to the above four components, we assign them into a category dubbed `in between' referring to \citet{2014A&A...562A..71B}.

\begin{figure}[!t]
\centering
\includegraphics[width=0.9\textwidth]{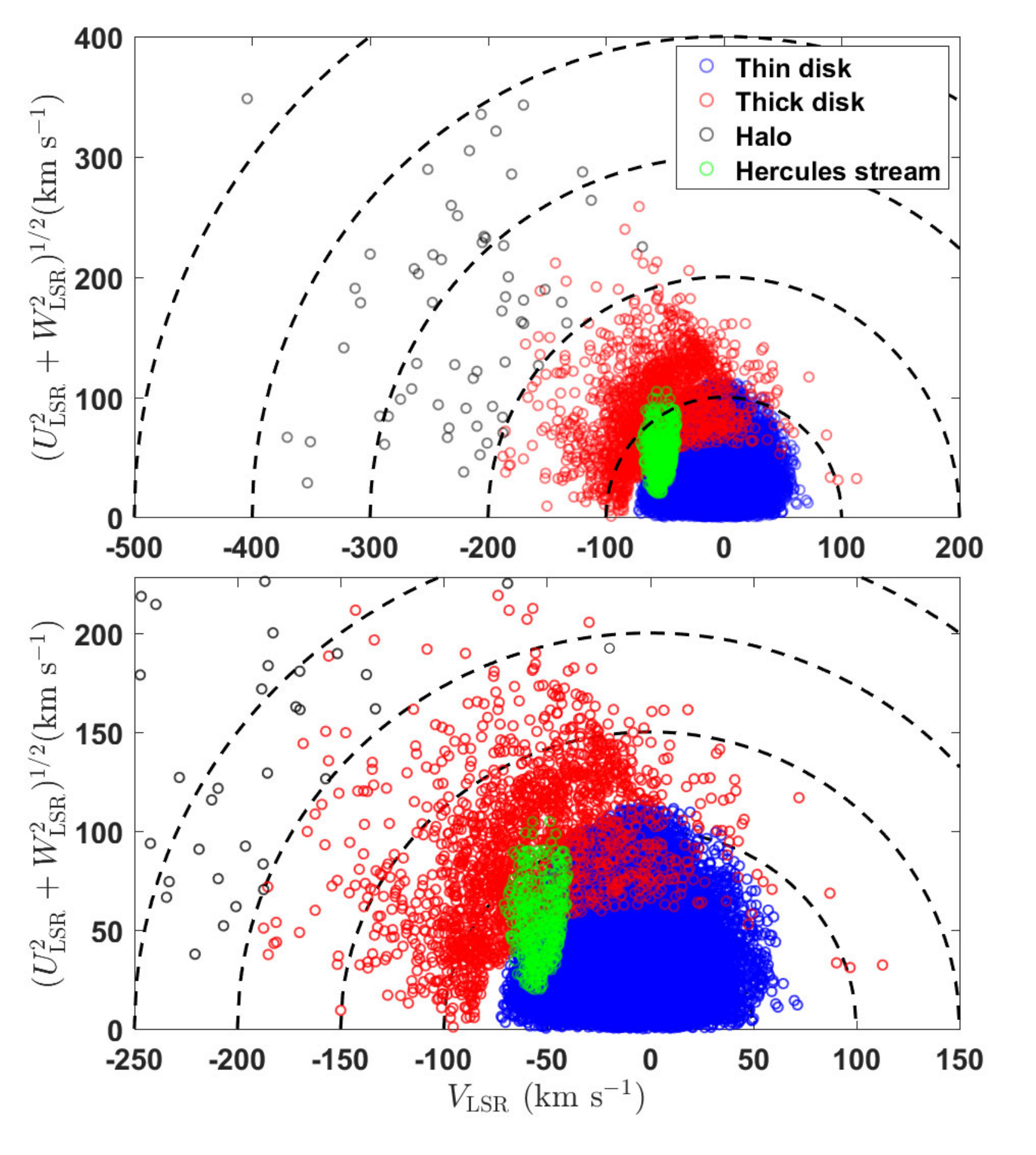}
\caption{The Toomre diagram of LAMOST-Gaia-Kepler stars for {\chendichang different Galactic components}. Top panel shows the full range of velocities while Bottom panel zooms in on the region where a majority of the sample is located.
The diagram are colour-coded to represent different memberships. 
Dotted lines show constant values of the total {\respondtoxiang Galactic} velocity $V_{\rm tot} = (U_{\rm LSR}^2+V_{\rm LSR}^2+W_{\rm LSR}^2)^{1/2}$, in steps of 100 and 50 $\rm km \ s^{-1}$ respectively, in the two plots.
\label{figGLKUVWpopu}}
\end{figure}

\begin{figure*}[!t]
\centering
\includegraphics[width=0.9\textwidth]{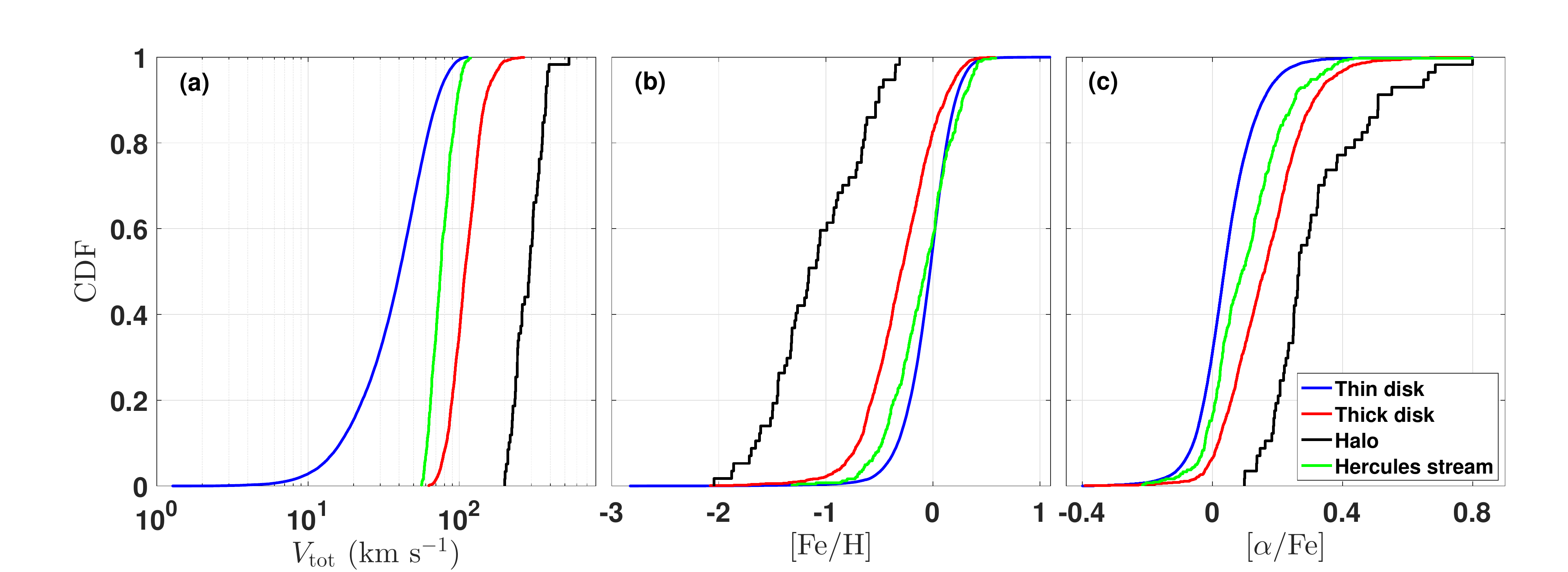}
\caption{The cumulative distributions of velocity $V_{\rm tot}$ (a), $\rm [Fe/H]$ (b) and $\rm [\alpha/Fe]$ (c) for LAMOST-Gaia-Kepler stars of different Galactic components, i.e., thin disk, thick disk, halo, and Hercules stream. 
\label{figVtotFealphaCDF}}
\end{figure*}

\begin{figure*}[!t]
\centering
\includegraphics[width=0.75\textwidth]{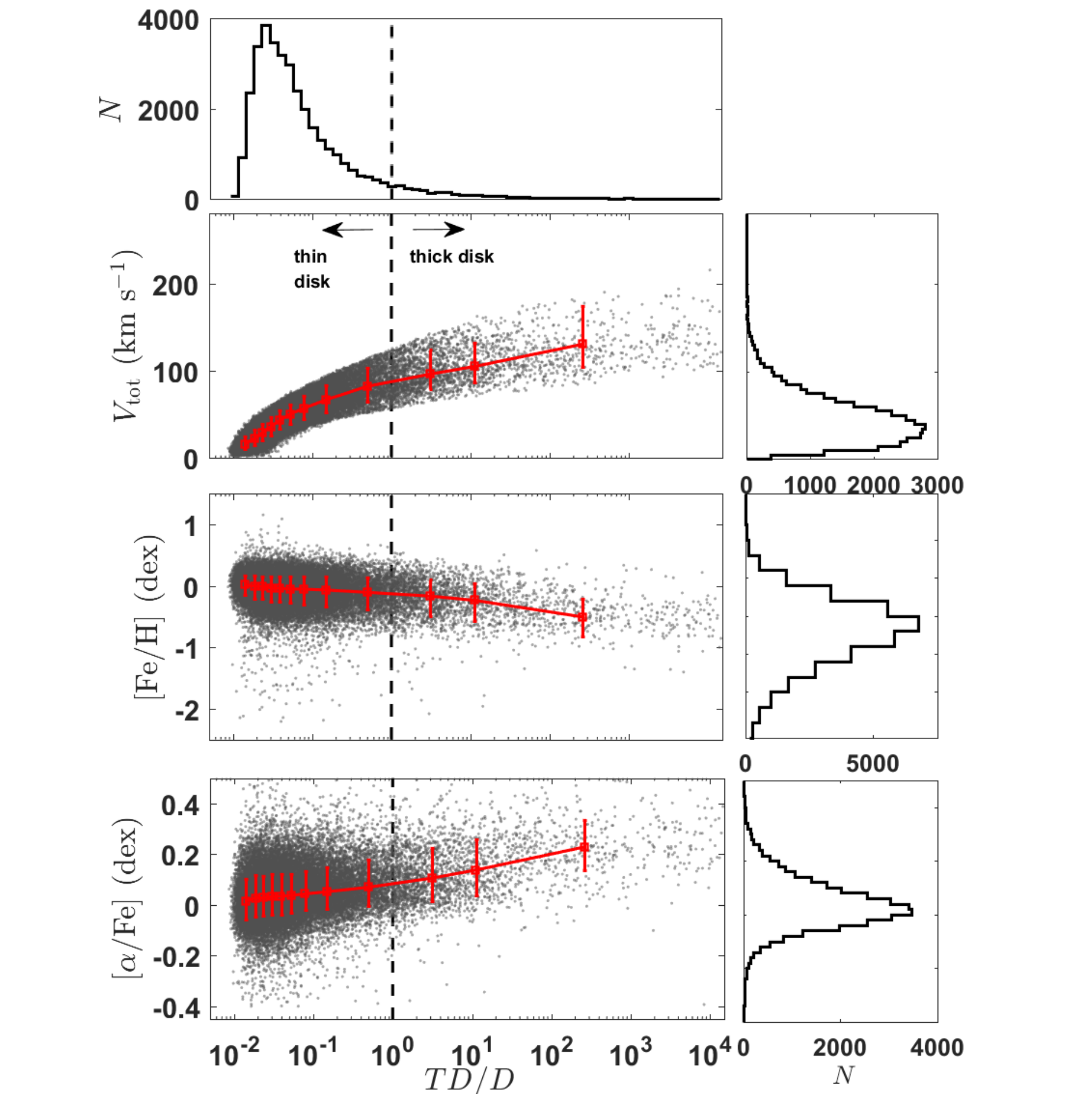}
\caption{Kinematic and chemical properties (from top to bottom: total velocity $V_{tot}$, abundances $\rm [Fe/H]$ and $\rm [\alpha/Fe]$) as functions of the relative probability between thick disk to thin disk (TD/D) for the LAMOST-Gaia-Kepler stars. 
Medians and 1 $\sigma$ dispersions are marked in the plots.
Histograms of $\rm V_{\rm tot}$, $\rm [Fe/H]$ and $\rm [\alpha/Fe]$ are shown in the right panels. Histogram of $TD/D$ is displayed in the {\chen topmost}.
The vertical dashed lines represents where $TD/D = 1$.
\label{figTDDVtotFealpha}}
\end{figure*}

\begin{table*}[!ht]
\centering
\caption{The catalogue of the LAMOST-Gaia-Kepler stellar sample}
{\footnotesize
\label{tab:starcatalog}
\begin{tabular}{p{1cm} p{2.5cm} p{1.25cm} p{1.25cm} p{8.75cm}} \hline
   Column & Name & Format & Units & description\\\hline \hline
\multicolumn{5}{c}{Parameters obtained from Gaia, LAMOST and NASA exoplanet archive (EA)} \\ \hline
  1 & Gaia\_ID & Long & & Unique Gaia source identifier \\
  2 & LAMOST\_ID & string & & LAMOST unique spectral ID\\
  3 & {Kepler}\_ID & integer & & Kepler Input Catalog (KIC) ID\\
  4 & {Gaia} RA & Double & deg & Barycentric right ascension\\  
  5 & {Gaia} Dec & Double & deg & Barycentric Declination\\ 
  6 & {Gaia} parallax & Double & mas & Absolute stellar parallax\\
  7 & {Gaia} e\_parallax & Double & mas & Standard error of parallax\\
  8 & {Gaia} pmra & Double & mas $\rm yr^{-1}$ & Proper motion in right ascension direction \\
  9 & {Gaia} e\_pmra & Double & mas $\rm yr^{-1}$ & Standard error of proper motion in right ascension direction\\
  10 & {Gaia} pmdec & Double & mas $\rm yr^{-1}$ &Proper motion in declination direction \\
  11 & {Gaia} e\_pmdec & Double & mas $\rm yr^{-1}$ & Standard error of proper motion in declination direction \\
  12 & {Gaia} G mag & Double & mag & \textit{Gaia} \textit{G} band apparent magnitude\\  
  13 & Kepler mag & Double & mag & \textit{Kepler} apparent magnitude \\ 
  14 & $T_{\rm eff}$ & Float & K & Effective temperature from LAMOST\\
  15 & e\_$T_{\rm eff}$ & Float & K & Error of effective temperature\\
  16 & $\log g$ & Float &  & Surface gravity from LAMOST\\
  17 & e\_$\log g$ & Float &  & Error of surface gravity from LAMOST\\
  18 & $\rm [Fe/H]$ & Float & dex &  Metallicity from LAMOST\\
  19 & e\_$\rm [Fe/H]$ & Float & dex & Error of metallicity \\
  20 & $\rm [\alpha/Fe]$ & Float & dex & $\alpha$ elements abundance from LAMOST \\ 
  21 & e\_$\rm [\alpha/Fe]$ & Float & dex & Error of $\alpha$ elements abundance\\ 
  22 & vr  &  Double & km $\rm s^{-1}$ &  Radial velocity from LAMOST\\          
  23 & e\_vr  &  Double & km $\rm s^{-1}$ & Error of radial velocity\\  
  24 & $N_{\rm p}$ & integer &  & Planet (candidate) multiplicity \\ \hline \hline
 \multicolumn{5}{c}{Parameters derived in this work} \\ \hline
  25 & $R$ & Double & kpc & Galactocentric Cylindrical radial distance \\  
  26 & $\theta$ & Double & deg & Galactocentric Cylindrical azimuth angle \\
  27 & $Z$ & Double & kpc & Galactocentric Cylindrical vertical height \\
  28 & $V_{R}$ & Double & km $\rm s^{-1}$ & Galactocentric Cylindrical $R$ velocities  \\ 
  29 & $V_{\theta}$ & Double & km $\rm s^{-1}$ & Galactocentric Cylindrical $\theta$ velocities  \\ 
  30 & $V_{Z}$ & Double & km $\rm s^{-1}$ & Galactocentric Cylindrical $Z$ velocities  \\ 
  31 & $U_{\rm LSR}$ & Double & km $\rm s^{-1}$ & Cartesian Galactocentric $x$ velocity to the LSR\\ 
  32 & e\_$U_{\rm LSR}$ & Double & km $\rm s^{-1}$ & error of Cartesian Galactocentric $x$ velocity to the LSR\\ 
  33 & $V_{\rm LSR}$ & Double & km $\rm s^{-1}$ & Cartesian Galactocentric $y$ velocity to the LSR\\ 
  34 & e\_$V_{\rm LSR}$ & Double & km $\rm s^{-1}$ & error of Cartesian Galactocentric $y$ velocity to the LSR\\ 
  35 & $W_{\rm LSR}$ & Double & km $\rm s^{-1}$ & Cartesian Galactocentric $z$ velocity to the LSR\\ 
  36 & e\_$W_{\rm LSR}$ & Double & km $\rm s^{-1}$ & error of Cartesian Galactocentric $z$ velocity to the LSR\\ 
  37 & $TD/D$ & Double & & thick disc to thin disc membership probability  \\
  38 & $TD/H$ & Double & & thick disc to halo membership probability  \\
  39 & $Herc/D$ & Double & & Hercules stream to thin disc membership probability  \\
  40 & $Herc/TD$ & Double & & Hercules stream to thick disk membership probability  \\ 
  41 &  Component & string & &  Classification of Galactic components \\ \hline \hline
\end{tabular}}
\end{table*}

\begin{table*}[!ht]
\centering
\renewcommand\arraystretch{1.25}
\caption{Numbers of stars ($N_s$) and planets ($N_p$) and kinematic and chemical properties of different Galactic components for the LAMOST-Gaia-Kepler sample.}
{\footnotesize
\label{tab:starproperty}
\begin{tabular}{l|cccccc} \hline
           &  {$N_{\rm s}$} & {$N_{\rm p}$} & $V_{\rm tot} \ \rm (kms^{-1})$ & $\rm [Fe/H]\  \rm (dex)$ & $\rm [\alpha/Fe] \ \rm (dex)$ & Age\_kin (Gyr) \\ \hline
    Thin disk & {\chen 31,218} & {\chen 955}  & $40.2^{+23.9}_{-19.9}$ &   $-0.03^{+0.18}_{-0.23}$ &  $0.04^{+0.09}_{-0.07}$ & $2.49^{+0.23}_{-0.19}$ \\
    Thick disk & 1,832 & 36 & $109.3^{+29.9}_{-22.5}$ & $-0.30^{+0.32}_{-0.32}$ &  $0.16^{+0.12}_{-0.12}$ & $9.83^{+1.16}_{-1.07}$\\
    Halo  &  59 & 0 & $291.3^{+64.9}_{-67.5}$ & $-1.16^{+0.53}_{-0.36}$ & $0.27^{+0.21}_{-0.08}$ & NA\\
    Hercules & 545 & 20 & $73.5^{+18.3}_{-12.3}$ & $-0.09^{+0.28}_{-0.32}$ &  $0.08^{+0.12}_{-0.09}$ & NA \\ \hline
 
\end{tabular}}
\end{table*} 

The numbers of stars and planets in {\chen different Galactic components} are summarized in Table \ref{tab:starproperty}.
As can be seen, about 87.1\% (31,218/{\chen 35,835}) of stars in our sample are in thin disk and about 5.1\% (1,832/{\chen 35,835}) stars are in thick disk. 
The fractions of halo and Hercules stream stars are about 0.16\% (59/{\chen 35,835}) and 1.5\% (545/{\chen 35,835}).
There are another $\sim$ 6.1\% (2,181/{\chen 35,835}) stars belonging to the `in between' category.

To display the distribution of velocities, in Figure \ref{figGLKUVWpopu}, we plot the Toomre diagram of the LAMOST-Gaia-Kepler stars. 
As can be seen, most stars with low velocities ($V_{\rm tot}\lesssim 70 \rm km\ s^{-1}$) are in the thin disk, while those with moderate velocities ($V_{\rm tot}\sim70-180 \rm km\ s^{-1}$) are mainly in thick disk. 
The velocities of halo stars are all larger than 200 $\rm km \ s^{-1}$.
For the Hercules stream, most of these stars have $V_{\rm LSR}$ around $-50 \ \pm \ 9 \ \rm km \ s^{-1}$ and $(U_{\rm LSR}^2+W_{\rm LSR}^2)^{1/2} \sim \rm 30-90 \ km \ s^{-1}$.
This is well consistent with the results derived by previous works \citep[e.g.][]{2003A&A...397L...1F,2013A&A...554A..44A,2014A&A...562A..71B,2017ApJ...845..101B}.

Then we compare the distributions of chemical abundances for different components in Figure \ref{figVtotFealphaCDF}.
As expected,  the thick disk stars are metal-poorer ($\sim 0.3$ dex) and $\alpha$-richer ($\sim 0.1$ dex) than the thin disk stars.
The Hercules stream stars have velocities and chemical abundances which are between those of thin and thick disk stars. 
For the halo, these stars have the highest Galactic velocities, poorest $\rm [Fe/H]$ and richest $\rm [\alpha/Fe]$.
In Figure \ref{figTDDVtotFealpha}, we plot the total velocity $V_{\rm tot}$, $\rm [Fe/H]$ and $\rm [\alpha/Fe]$ as a function of $TD/D$. As can be seen, with the increasing of $TD/D$, $V_{\rm tot}$ and $\rm [\alpha/Fe]$ increase, while $\rm [Fe/H]$ decreases.
Because there seems no clear trend between velocity dispersions and ages for stars in the halo and Hercules stream \citep{2021ApJ...909..115C}, here we only compare the age distributions for stars in the Galactic disk.
The typical ages are $2.43^{+0.32}_{-0.19}$ Gyr, $9.83^{+1.30}_{-1.02}$ Gyr for thin and thick disk respectively.
The Numbers of stars and planets, kinematic properties and chemical abundances of different Galactic components are summarized in Table \ref{tab:starproperty}. 

For the sake of completeness, we also add the stellar parameters that used during the process of our kinematic characterization (e.g., parallax, proper motion and RV) and other basic stellar parameters (e.g., $T_{\rm eff}$, $\log g$, $\rm [Fe/H]$ and $\rm [\alpha/Fe]$) into the catalog.
In what follows, we conduct some further analyses and discussions on this catalog.



\section{Further Catalog Analyses and Discussions}
\label{sec.cat.analy}
In this section, based on the LAMOST-Gaia-Kepler catalog, we perform some tests and comparisons to verify the reliability of the kinematic age (section \ref{sec.cat.analy.vetify}), explore the evolution of stellar magnetic activity with kinematic ages (section \ref{sec.cat.stellaractivity}), and analyze the kinematic properties of Kepler {\chendichang planet candidate host} stars  (section \ref{sec.res.planetvsnonplanethost}).  

\begin{figure*}[!t]
\centering
\includegraphics[width=\textwidth]{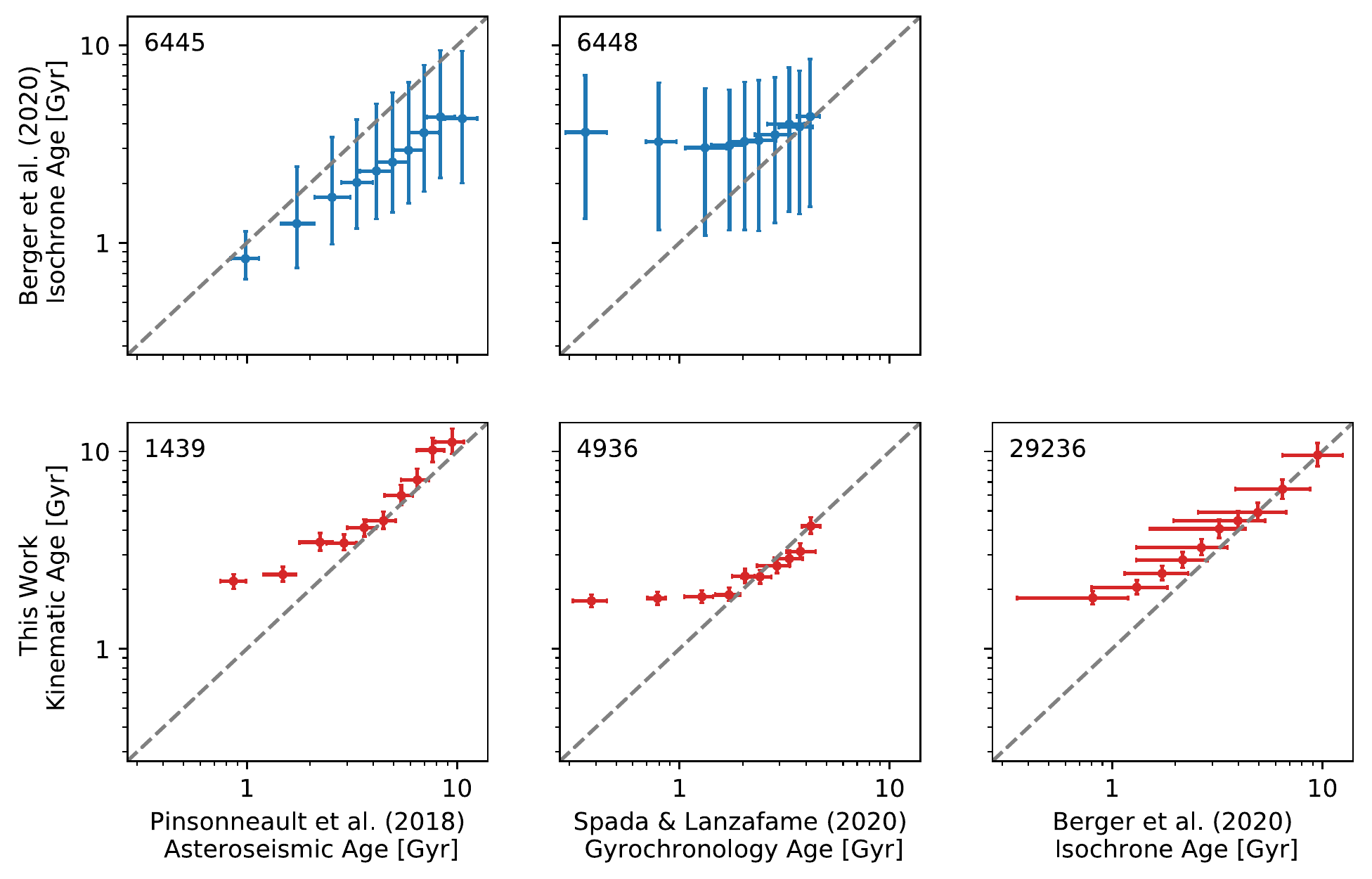}
\caption{Comparison among ages derived by various methods. The x-axes for each columns from left to the right are asteroseismic age from \citet{Pinsonneault.2018ApJS..239...32P}, gyrochronology age derived from method provided by \citet{Spada.2020A&A...636A..76S}, and isochrone age from \citet{2020AJ....159..280B}, y-axes for each row are isochrone age and kinematic age derived in this work, respectively. In every subplot, we compare the ages from the corresponding x and y samples, the dots and errorbars are the median age and median value of relative error, except kinematic ages and uncertainties are calculated by method detailed in section \ref{sec.kine.age.err}. We label the size of sample in the upper left corner, and the gray dash line shows where x equals to y. 
\label{fig_age}}
\end{figure*}


\subsection{Verifying the Kinematic Age}
\label{sec.cat.analy.vetify}
In order to verify the reliability of the kinematic age, we compare it with ages derived from other methods, such as asteroseismology, gyrochronology, and isochrone fitting in this subsection.

With seismic parameters $\Delta\nu$ and $\nu_{\text{max}}$ derived from Kepler asteroseismic data, and spectroscopic parameters from Apache Point Observatory Galactic Evolution Experiment (APOGEE) project \citep{2017AJ....154...94M}, \citet{Pinsonneault.2018ApJS..239...32P} provided asteroseismic ages for 6,676 evolved stars. 

Gyrochronology can estimate the age of a main sequence star based on its rotation period \citep{Barnes.2010ApJ...722..222B,Barnes.2010ApJ...721..675B}. We use rotation data for Kepler stars from \citet{McQuillan.2013MNRAS.432.1203M,McQuillan.2013ApJ...775L..11M,McQuillan.2014ApJS..211...24M}, and derive gyrochronology ages for 5,851 stars by adopting the method provided by \citet{Spada.2020A&A...636A..76S}.

Fitting isochrone grid provided by the latest MIST models \citep{Dotter.2016ApJS..222....8D,Choi.2016ApJ...823..102C,Paxton.2011ApJS..192....3P,Paxton.2013ApJS..208....4P,Paxton.2015ApJS..220...15P,Paxton.2018ApJS..234...34P},  \cite{2020AJ....159..280B} obtained isotropic ages for 186,301 Kepler stars with a typical uncertainty of $\sim 56\%$. Here, we only consider stars with relatively reliable isochrone ages by applying the selection criteria as suggested by \citet{2020AJ....159..280B}, i.e., stars with GOF$>$0.99, having TAMS less than 20 Gyr, and metallicity measured by spectroscopic method. We also remove stars with isochrone age longer than 14 Gyr.

In Figure \ref{fig_age}, we compare kinematic age inferred from this work and ages derived with other methods to each other. 
Subplots in each column and row share the same x-axis and y-axis, respectively. We label the methods to derive age and the relevant literatures in the corresponding coordinates. 
In each subplot, we cross-match the two samples shown in the x and y axes, and print the number of stars in the matched sample in the left upper corner. 
We sort each sample {\chen in ascending order} according to the x-axis, and divide them into 10 bins with approximately equal size. 
For asteroseismic, gyrochronology, and isochrone ages, the dots and errorbars shown in Figure \ref{fig_age} are median ages and median values of relative error.
For the kinematic age, we calculate the age and uncertainty with methods mentioned in section \ref{sec.kine.age.err}. 

{\chen S}ince ages derived from asteroseismology are mainly for giant stars and gyrochronology applies only to main sequence stars, there is no common star between these two samples and thus comparison has never been made between gyrochronology age and asteroseismic age yet.
In the {\chen upper} panels, isochrone age is compared with asteroseismic age and gyrochronology age. 
Isochrone age exhibits systematic deviation from asteroseismic age and apparent discrepancy with gyrochronology age for stars younger than 3 Gyrs, though in most of cases they are generally consistent with each other due to the large uncertainty of isochrone age.
In the lower panels, we show comparisons between kinematic age and other ages.
In general, kinematic age matches well with ages derived from asteroseismic, gyrochronology, and isochrone, with {\chen relative large differences only in the young end $(<$2 Gyrs for asteroseismic age, $<$3 Gyrs for gyrochronology and isochrone ages$)$. }
Such a discrepancy is not unexpected because the stellar velocity dispersion at early days might be dominated by the initial condition rather than dynamical evolution and thus age derived from velocity dispersion would be overestimated.

\subsection{Stellar Magnetic Activity Evolves with Kinematic Age}
\label{sec.cat.stellaractivity}
Stellar properties, such as rotation period and stellar activity, are indicators of stellar age. We explore the connections between them and kinematic age in this subsection.

\begin{figure}[!t]
\centering
\includegraphics[width=0.75\textwidth]{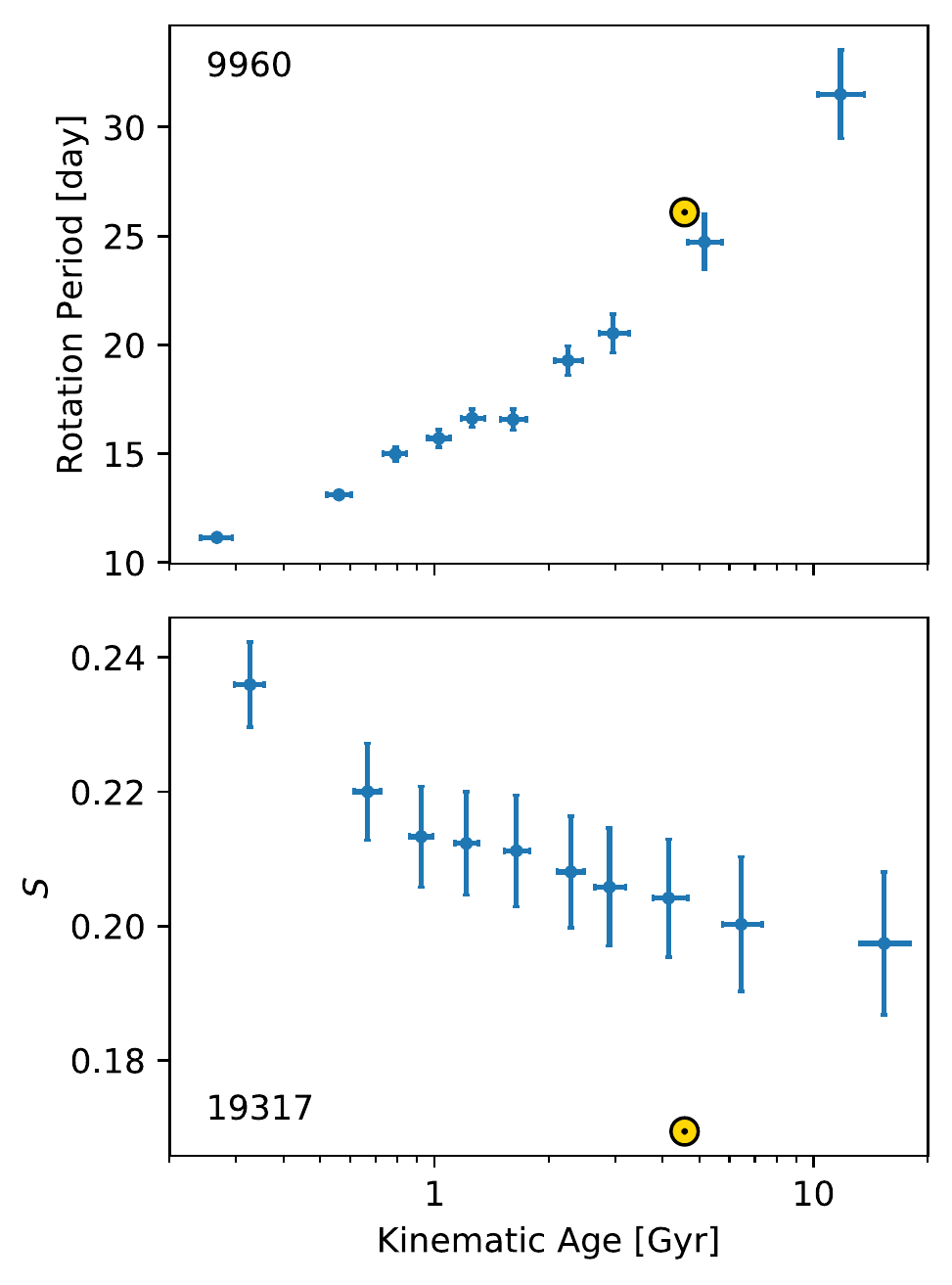}
\caption{Stellar rotation period and magnetic activity ($S$-index) as functions of kinematic age. In the upper panel, the dots present the kinematic age and stellar rotation period. The x-axis errorbars are derived from section \ref{sec.kine.age.err}, and errorbars for the y-axis are {\chen median of rotation period and median value of relative error}. The comparison between kinematic age and $S$-index is shown in the lower panel, with dots and errorbars derived with the same method. In both of the panels, the number of the sample size are shown in the upper/lower left corner, the yellow dots present the age, rotation period, and $S$-index of the Sun.
\label{fig_rot_rhk}}
\end{figure}

We match our LAMOST-Gaia-Kepler sample with rotation periods from \cite{McQuillan.2013MNRAS.432.1203M,McQuillan.2013ApJ...775L..11M,McQuillan.2014ApJS..211...24M} by KIC, obtaining 10,548 common stars.
We sort the sample according to the $TD/D$ value, divide them into 10 groups with approximately equal size and calculate kinematic ages and uncertainties for each group using the method mentioned in section \ref{sec.kine.age.err}.
For each bin, we calculated the {\chen median rotation period and median value of relative error}, and show them with dots and errobars in the upper panel of Figure \ref{fig_rot_rhk}.
As kinematic age increases, rotation period becomes longer, which is in agreement with the gyrochronology theory \citep{Barnes.2010ApJ...721..675B,Barnes.2010ApJ...722..222B}.
For comparison, we show the age (4.57 Gyr) and rotation period (16.09 days) of the Sun with a yellow dot.
As can be seen, the Sun is a typical star which fits well to the kinematic age--rotation period trend.

In our LAMOST--Gaia--Kepler sample, 20,417 stars have the $S$-index meausrements based on the LAMOST spectra \citep{Zhang.2020ApJS..247....9Z}.
We group the sample into 10 subgroups.
For each bin, we calculate the median value of $S$-index and the uncertainty is set as {\chen median value of relative error}, which are plotted as dots and errobars in the upper panel of Figure \ref{fig_rot_rhk}. 
As stars aging, stellar activity reduces and the median value of $S$-index decreases as expected.
For $S$-index of the Sun, we adopt the mean value $<S>$=0.1694 suggested by \citet{Egeland.2017ApJ...835...25E}, and plot it in the figure with a yellow dot.
As can be seen, the Sun is exceptionally quiet as compared to stars in the LAMOST--Gaia--Kepler sample, confirming the results of previous studies \citep[e.g.,][]{Reinhold.2020Sci...368..518R}.





\begin{figure}[!t]
\centering
\includegraphics[width=0.75\textwidth]{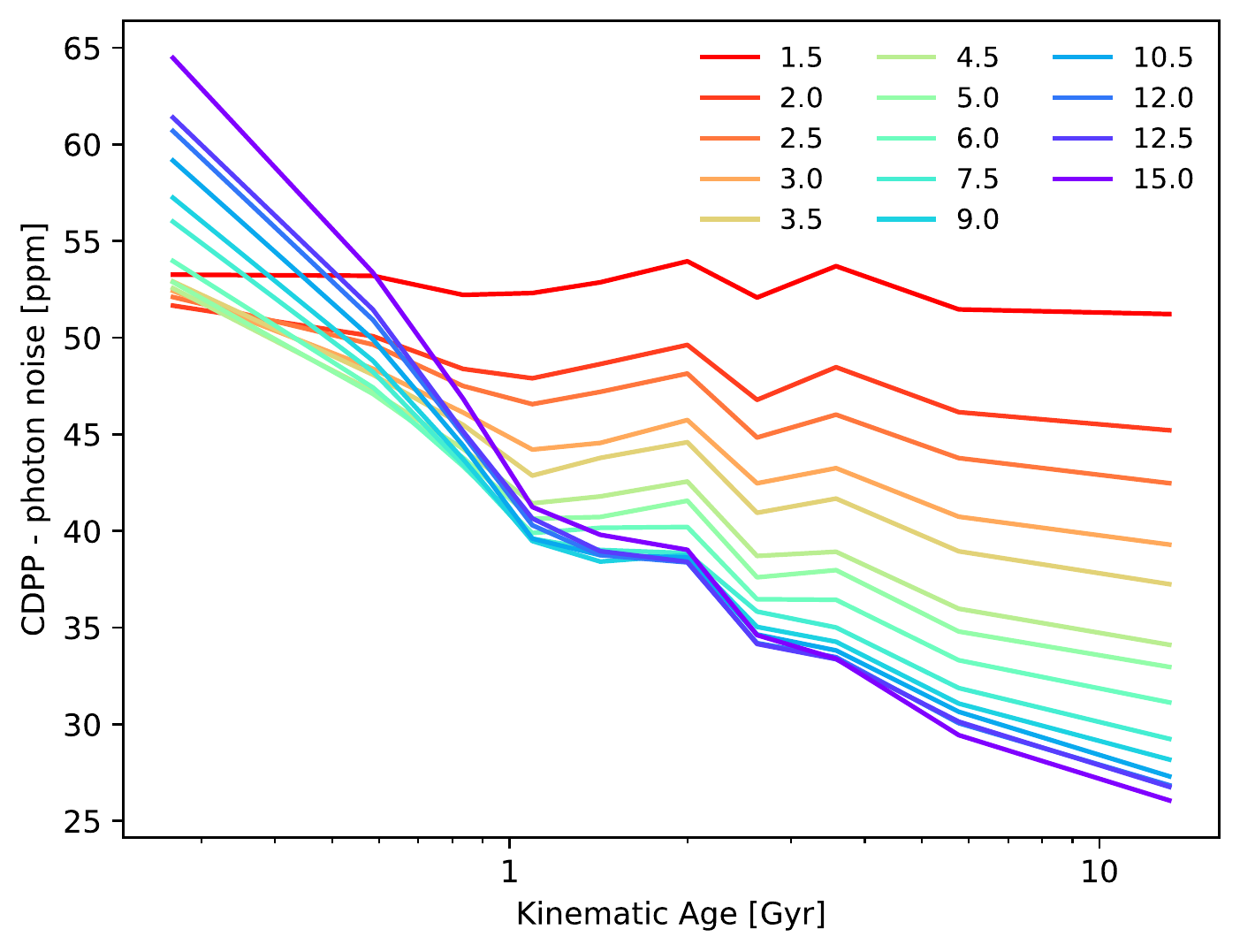}
\caption{Comparison between kinematic age and {\chen CDPP$-$photon} noise. Each line presents CDPP for a certain time scale, and shown in the legend.
\label{fig_cdpp}}
\end{figure}

\begin{figure*}[!t]
\centering
\includegraphics[width=0.95\textwidth]{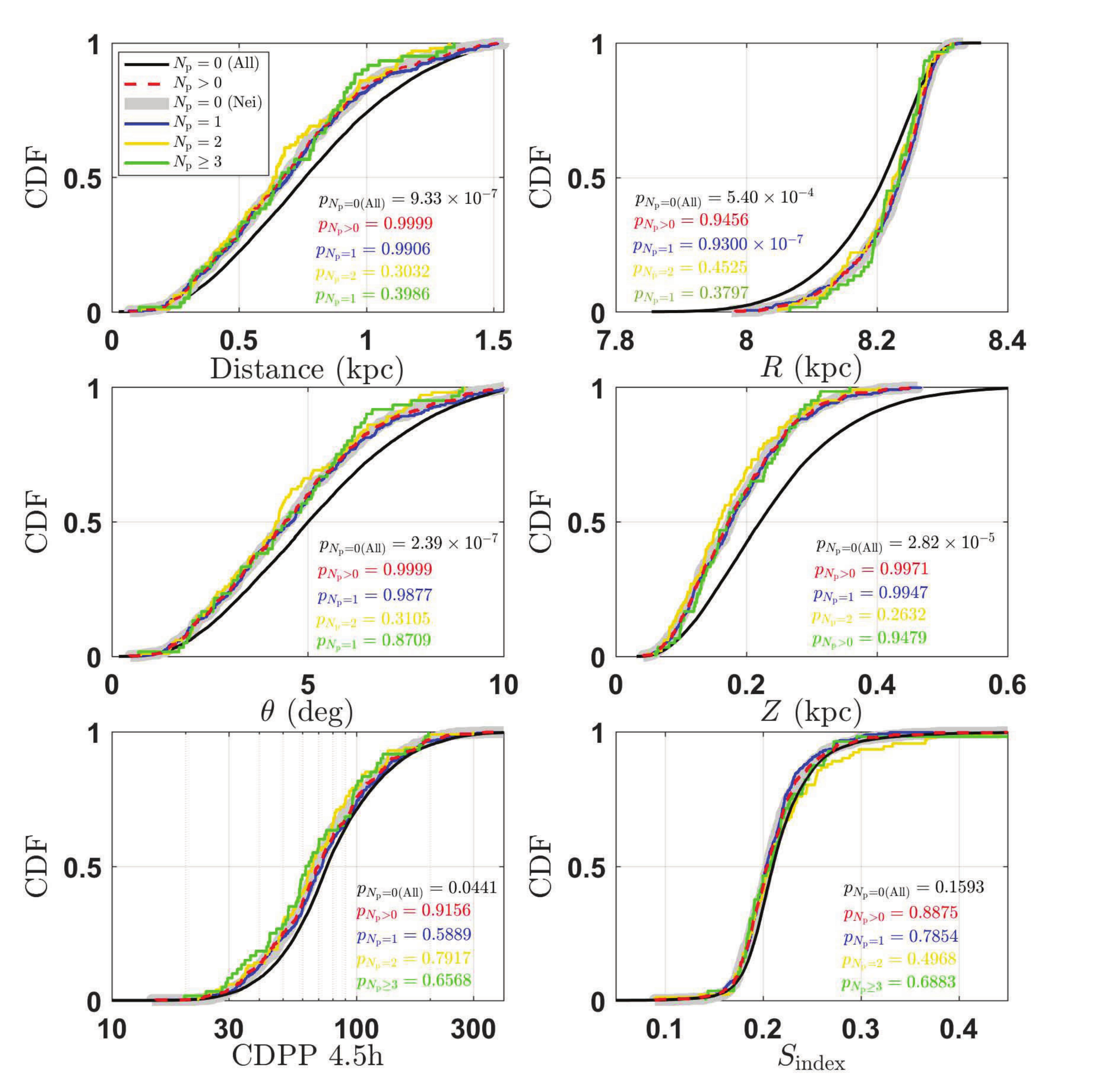}
\caption{\chen The cumulative distributions of distance to the Sun, Galactic coordinates ($R$, $\theta$, $Z$), CDPP 4.5h and $S_{\rm index}$ for Kepler {\chendichang planet candidate host} stars (dashed red lines, labelled as $N_{\rm p}>0$) and all stars without Kepler planets (solid black lines, labelled as $N_{\rm p}=0$ (All)). 
The distributions of the subsamples divided by the numbers of transit planets are plotted in different colours: blue for $N_{\rm p} =1$, yellow for $N_{\rm p} =2$ and green for $N_{\rm p} \ge 3$.
The wide grey lines represent the distributions of the selected nearest neighbors for every Kepler {\chendichang planet candidate host} star from stars without Kepler planets (labelled as $N_{\rm p=0}$ (Nei)).
In each panel, the $p$ denotes the $p-$value of the two sample KS test for the distributions of the other samples (the subscripts same as the labels) comparing to the neighbor stars without Kepler planets ($N_{\rm p} = 0$ (Nei)). 
\label{figCDFSpatial}}
\end{figure*}

Stellar magnetic activity can also be revealed by photometry.
In Figure \ref{fig_cdpp}, we investigate how the photometric noise levels of stars change with kinematic age. 
For every target, Kepler DR 25 provides Combined Differential Photometric Precision  \citep[CDPP,][]{Christiansen.2012PASP..124.1279C} for 14 different time scales, which characterises the noise level in Kepler lightcurves.
In order to minimize other parameters that may affect CDPP other than age, such as evolve state, magnitude, and spectral type, we restrict out sample to stars with $\log g>4$, Kepler magnitude (kepmag) lower than 14, and effective temperature between 4,700K to 6,500K, which contains 14,372 stars.
Since CDPP is partly contributed by photon noise, following the method of \citet{Bastien.2013Natur.500..427B}, we subtract photon noise in quadrature from CDPP, i.e., {\chen CDPP$-$photon} noise.
Again, we group the star sample into 10 bins then calculate their corresponding kinematic ages and the median value of {\chen CDPP$-$photon} noise.
We show the {\chen CDPP$-$photon} -- Age noise relationship in Figure \ref{fig_cdpp}, each line with a different color presents CDPP with a different duration.
For longer duration, {\chen CDPP$-$photon} noise decrease as stars aging, 15 hours {\chen CDPP$-$photon} noise drops from 65 ppm to 25 ppm as kinematic age grows from 0.3 Gyr to 12 Gyr.
\citet{Gilliland.2011ApJS..197....6G} modelled the stellar noise level as a function of age and found that the total noise on 6.5 hours timescale which is dominated by stellar activity decreases as stars aging.
Our result support their model from the aspect of observation.
For shorter duration, the declining trend between {\chen CDPP$-$photon} noise and kinematic age is weaker.
This is not unexpected because shorter timescale noises could be contributed by granulation which generally increases with age, and thus compensating the decline trend.
{\chen For main sequence stars, the typical timescale of granulation {\chendichang is on order of} a few hundreds seconds \citep{Gilliland.2011ApJS..197....6G,Kallinger.2014A&A...570A..41K}, the effect of granulation will be more evident on CDPP with shorter range, such as 1.5 hours.}


\begin{table*}[!ht]
\centering
\renewcommand\arraystretch{1.28}
\caption{Stellar properties of Kepler {\chendichang planet candidate host} stars and stars without Kepler planets.
}
{\footnotesize
\label{tab:starpropertywpnp}
\begin{tabular}{l|cc|cccc} \hline
                &  All Stars without  & neighbor stars without & \multicolumn{4}{c}{---------------------~~Kepler {\chendichang planet candidate host} stars~~---------------------}    \\ 
                & Kepler planets ($N_{\rm p}=0$) & Kepler planets ($N_{\rm p}=0$) & All & $N_{\rm p}=1$ & $N_{\rm p}=2$ & $N_{\rm p} \ge 3$  \\ \hline
    {$M_* \ (M_{\odot})$} &  {$1.06^{+0.21}_{-0.18}$}  & {\chen $1.04^{+0.20}_{-0.19}$} &  {$1.03^{+0.21}_{-0.16}$} & {$1.03^{+0.20}_{-0.16}$} & {$1.03^{+0.18}_{-0.17}$} &  {$1.05^{+0.22}_{-0.17}$}  \\ 
    {$R_* \ (R_{\odot})$} &   {$1.20^{+0.41}_{-0.31}$}  & {\chen $1.15^{+0.38}_{-0.30}$} & {$1.10^{+0.40}_{-0.23}$} & {$1.09^{+0.39}_{-0.23}$} & {$1.12^{+0.38}_{-0.27}$} &  {$1.13^{+0.38}_{-0.26}$}  \\ 
    {$\rm [Fe/H] \ \rm (dex)$} & {$-0.02^{+0.19}_{-0.26}$}  & {\chen $-0.01^{+0.19}_{-0.27}$} & {$0.00^{+0.18}_{-0.21}$} & {$-0.01^{+0.19}_{-0.21}$} & {$0.01^{+0.13}_{-0.23}$} &  {$-0.02^{+0.17}_{-0.18}$}  \\ 
    {$\rm [\alpha/Fe] \ \rm (dex)$} & {$0.05^{+0.10}_{-0.08}$}  & {\chen $0.05^{+0.09}_{-0.07}$} & {$0.04^{+0.09}_{-0.08}$} & {$0.04^{+0.09}_{-0.08}$} & {$0.03^{+0.09}_{-0.07}$} &  {$0.05^{+0.07}_{-0.07}$}  \\ 
    \multirow{2}{*}{$F_{\rm D}$} &{\chen $ 86.9^{+0.23}_{-0.23}\%$ } &{\chen $86.8^{+1.4}_{-1.4}\%$} & {\chen $88.8^{+1.3}_{-1.3}\%$}  & {\chen $88.6^{+1.6}_{-1.6}\%$}  & $90.0^{+3.2}_{-3.1}\%$  & $91.7^{+3.6}_{-3.5}\%$    \\
          & {\chen (18079/20795)} & {\chen (480/553)} & {\chen (502/563)} & {\chen (357/403)} & (90/100) & (55/60)  \\
    \multirow{2}{*}{$F_{\rm TD}$} & $5.2^{+0.2}_{-0.2}\%$  & {\chen $5.8^{+1.0}_{-1.0}\%$} & {\chen $4.4^{+0.9}_{-0.9}\%$}  & {\chen $5.5^{+1.2}_{-1.2}\%$}  & $3.0^{+1.7}_{-1.9}\%$  & $0^{+3.1}_{-0}\%$   \\
    & {\chen (1088/20795)} & {\chen (32/553)} & {\chen (25/563)} & {\chen (22/403)} & (3/100) & (0/60)  \\
    \multirow{2}{*}{$F_{\rm H}$} & $0.17^{+0.02}_{-0.02}\%$  & {\chen $0.36^{+0.48}_{-0.23}\%$} & {\chen $0^{+0.3}_{-0}\%$}  & {\chen $0^{+0.5}_{-0}\%$}  & $0^{+1.8}_{-0}\%$  & $0^{+3.1}_{-0}\%$   \\
    & {\chen (35/20795)} & {\chen (2/553)} & {\chen (0/563)} & {\chen (0/403)} & (0/100) & (0/60)  \\
    \multirow{2}{*}{$F_{\rm Herc}$} & $1.6^{+0.08}_{-0.08}\%$   & {\chen $0.90^{+0.61}_{-0.39}\%$} & {\chen $1.8^{+0.6}_{-0.5}\%$}  & {\chen $1.5^{+0.6}_{-0.5}\%$}  & $3.0^{+1.7}_{-1.9}\%$  & $1.7^{+1.72}_{-1.4}\%$  \\
 & {\chen (329/20795)} & {\chen (5/553)} & {\chen (10/563)} & {\chen (6/403)} & (3/100) & (1/60)  \\
    {$V_{\rm tot} \ (\rm km \ s^{-1})$} & {\chen $44.06^{+31.24}_{-22.32}$}  & {\chen $43.05^{+32.04}_{-21.13}$} & {$36.37^{+33.70}_{-16.36}$} & {$37.89^{+33.69}_{-17.88}$} & {$32.30^{+33.64}_{-13.21}$} &  {$29.36^{+30.87}_{-10.88}$}  \\ 
    {$\sigma_{\rm tot} \ (\rm km \ s^{-1})$} &  {\chen $49.07^{+0.84}_{-0.73}$}  & {\chen $49.99^{+1.25}_{-0.91}$} & {\chen $ 45.65^{+1.14}_{-0.77}$} & {\chen $46.92^{+1.20}_{-1.21}$} & {$42.48^{+1.64}_{-1.32}$} &  {$41.70^{+2.11}_{-1.87}$}  \\ 
    {$\rm Age\_{kin} \ (\rm Gyr)$} & {\chen $4.28^{+0.55}_{-0.35}$}  & {\chen $ 4.49^{+0.60}_{-0.36}$} & 
    {\chen $3.57^{+0.43}_{-0.27}$} & {\chen $3.82^{+0.50}_{-0.28}$} & {$2.98^{+0.36}_{-0.20}$} & {$2.84^{+0.34}_{-0.29}$}  \\
    \hline
\end{tabular}}
\flushleft
{\scriptsize
$F$ represents the number fraction of different components: thin disk (D), thick disk (TD), halo (H) and Hercules stream (Herc). 
\\}
\end{table*}   


\begin{figure*}[!t]
\centering
\includegraphics[width=\textwidth]{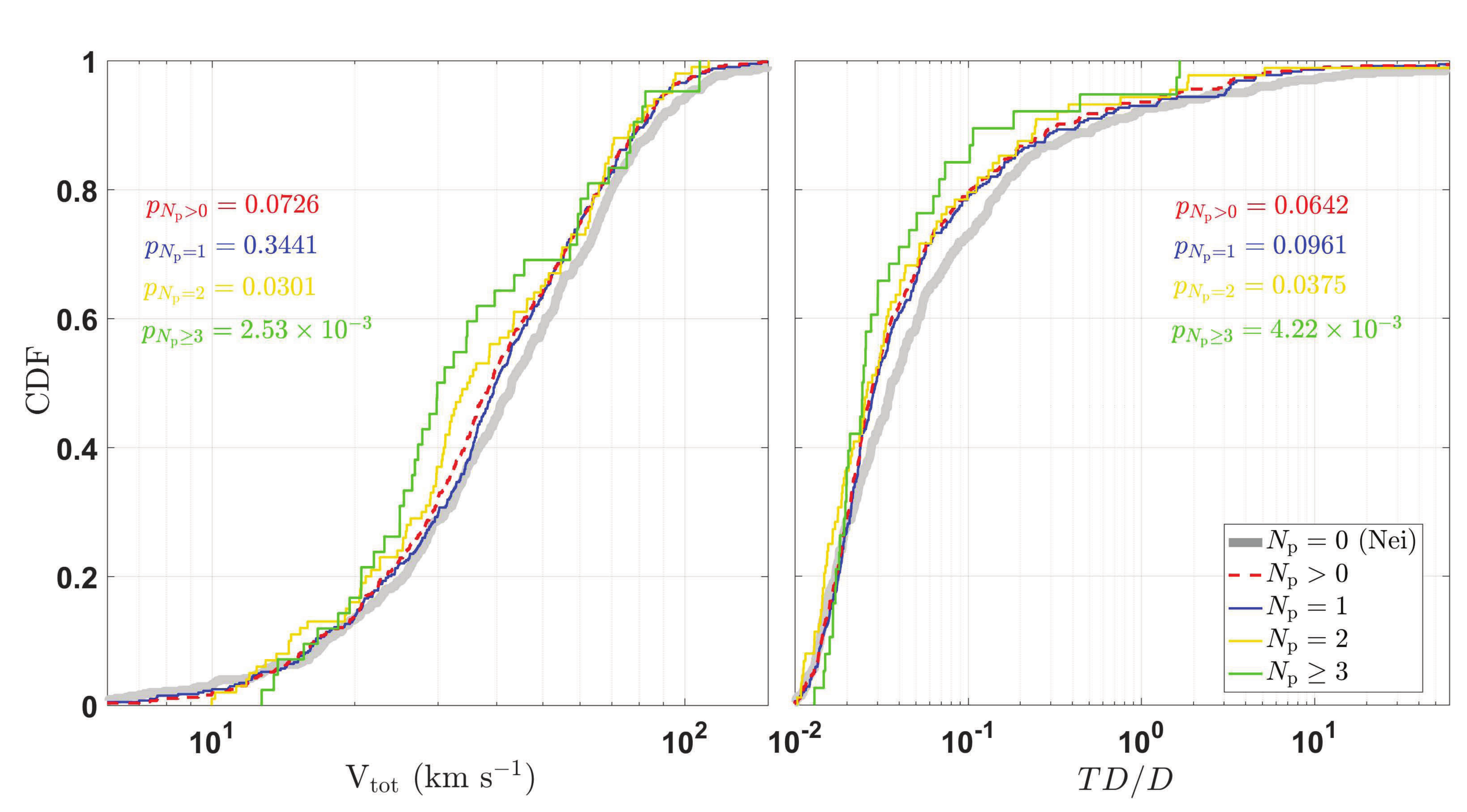}
\caption{\chendichang The cumulative distribution of Galactic velocity, $V_{\rm tot}$ (left panel) and the relative probabilities of the  thick-disk-to-thin-disk, $TD/D$ (right panel)  for all {\chendichang planet candidate host} (dashed red lines) 1 transiting {\chendichang planet candidate host}s (blue lines), 2 transiting {\chendichang planet candidate host}s (yellow lines), $3+$ transiting {\chendichang planet candidate host}s (green lines) and neighbor stars without Kepler planets (wide grey lines, labelled as $N_{\rm p}=0$ (Nei)). 
In each panel, the $p$ denotes the $p-$value of the two sample KS test for the distributions of the other samples (the subscripts same as the labels) comparing to the neighbor stars without Kepler planets ($N_{\rm p} = 0$ (Nei)).
\label{figCDFvtotTDD}}
\end{figure*}

\subsection{Kinematic properties of Kepler {\chendichang planet candidate host} Stars}
\label{sec.res.planetvsnonplanethost}
In this section, we explore the differences between the kinematic properties of Kepler {\chendichang planet candidate host} stars and those of other stars  without Kepler planets, which are meaningful to study the formation and evolution environment of planetary systems.
To avoid the influence of the stellar evolutionary stage and spectral type, here we only compare the distribution of Solar-type stars, which are the bulk of LAMOST-Gaia-Kepler sample. 
The selection criteria are taken as an effective temperature $T_{\rm eff}$ in the range of $4700–6500$ K and a stellar surface gravity $\log g>4.0$. 
Finally, we are left with {\chen 563} Kepler {\chendichang planet candidate host}s and {\chen 20,795} stars without Kepler planets.

\subsubsection{Spatial distribution}
\label{sec.res.planethost.space}
Here we compare the distribution of spatial position in Figure \ref{figCDFSpatial}. As can be seen, 
the stars without Kepler planets (solid black lines) have a wider distribution in the distance to the Sun, Galactic radius ($R$), azimuth angle ($\theta$) and height ($Z$) than that of Kepler {\chendichang planet candidate host} stars (dashed red lines). 
We did Kolmogorov–Smirnov (KS) tests between the distance to the Sun, Galactic coordinates ($R$, $\theta$, $Z$) of the Kepler {\chendichang planet candidate host} stars and those of stars without Kepler planets. The resulted $p$ values are all smaller than 0.003, demonstrating that there are significant differences between their spatial positions.
{\chendichang As shown in Figure \ref{fig_rot_rhk} and \ref{fig_cdpp},  the stellar activity (and therefore noise properties) are correlated to their kinematics. This may affect the detectability of planets in multi-planet systems, e.g., by creating a bias against smaller planets around active/noisy stars (typically corresponding to younger stars)}
Thus in the following comparison in stellar kinematic properties, to avoid the bias caused by spatial distribution and stellar activity/noise, we construct a control sample by adopting the NearestNeighbors function in scikit-learn \citep{10.5555/1953048.2078195} to select the nearest neighbors for every host star from stars without Kepler planets in spatial distribution (i.e. $R$, $\theta$, and $Z$), CDPP and $S_{\rm index}$.
Here we select the CDPP of 4.5 hour as the transit duration are 4.3 hours by taking the median values of stellar mass (1.03 $M_\odot$), radius (1.10 $R_\odot$) and planetary period (10.37 days) in our sample.
In the case that a star without planets is selected as the nearest neighbors of multiple times, we exclude duplicate stars.
The neighbor stellar sample without Kepler planets contains {\chen 551} stars.
{\chen As can be seen in Figure \ref{figCDFSpatial}, the distributions of the neighbor stars without Kepler planets (wide grey lines) are very similar to that of Kepler {\chendichang planet candidate host} stars (dashed red lines). 
We also did KS tests between their spatial distributions and stellar activity/noise. The p-values are 0.9999, 0.9456, 0.9877, 0.9971, 0.9156 and 0.8875 for distance, $R$, $\theta$, $Z$, CDPP 4.5h and $S_{\rm index}$ respectively. Such high p-values demonstrate the spatial distributions and stellar activity/noise of the {\chendichang planet candidate host} stars and neighbor stars without Kepler planets are statistically indistinguishable.}
Therefore, the selected neighbor stars form a reliable control sample to minimize the spatial and detection bias.

In the following discussions,  we divided the Kepler {\chendichang planet candidate host} sample into three subsmaples according to the number of transiting planets (tranets) in each system: $N_{\rm p}=1$ ({\chen 403} stars), $N_{\rm p}=2$ (100 stars), $N_{\rm p} \ge 3$ (60 stars).
The p-values of KS test between the distribution of the distance and Galactic coordinates for the neighbor stellar sample without Kepler planets and the three subsamples (blue, yellow, red lines in Figure \ref{figCDFSpatial}) are all larger than 0.25, which prove that their spatial distribution are statistically indistinguishable. 
{\chen 
In Table \ref{tab:starpropertywpnp}, we summarized the typical value of physical properties.
The uncertainties are taken as the 50$\pm$34.1 percentiles of the corresponding distributions. 
As can be seen, their typical values in mass, radius, $\rm [Fe/H]$  and $\rm [\alpha/Fe]$ are indistinguishable within the 1--$\sigma$ uncertainties.
We also make KS test and resulted p-values are all larger than 0.1, demonstrating that there is no significant difference in the distributions of mass, radius, $\rm [Fe/H]$  and $\rm [\alpha/Fe]$.}

So far, we have found that Kepler {\chendichang planet candidate host}s have a narrower range of Galactic positions ($R$, $\theta$, $Z$) and shorter distance as compared to the general Kepler field stars.
This is likely to be caused by the observational bias, namely, planets are more easily found around stars closer to the Earth.
{\chen To remove this spatial and detection bias, we constructed a control sample by selecting the neighbor stars of {\chendichang planet candidate host}s. 
As we will show below, although {\chendichang planet candidate host}s have similarities with the control sample in  Galactic position, mass, radius, chemical abundance}, they (especially those with large planet multiplicity, e.g., $N_{\rm p} \ge 3$) do differ significantly in kinematic properties.

\begin{figure}[!t]
\centering
\includegraphics[width=0.75\textwidth]{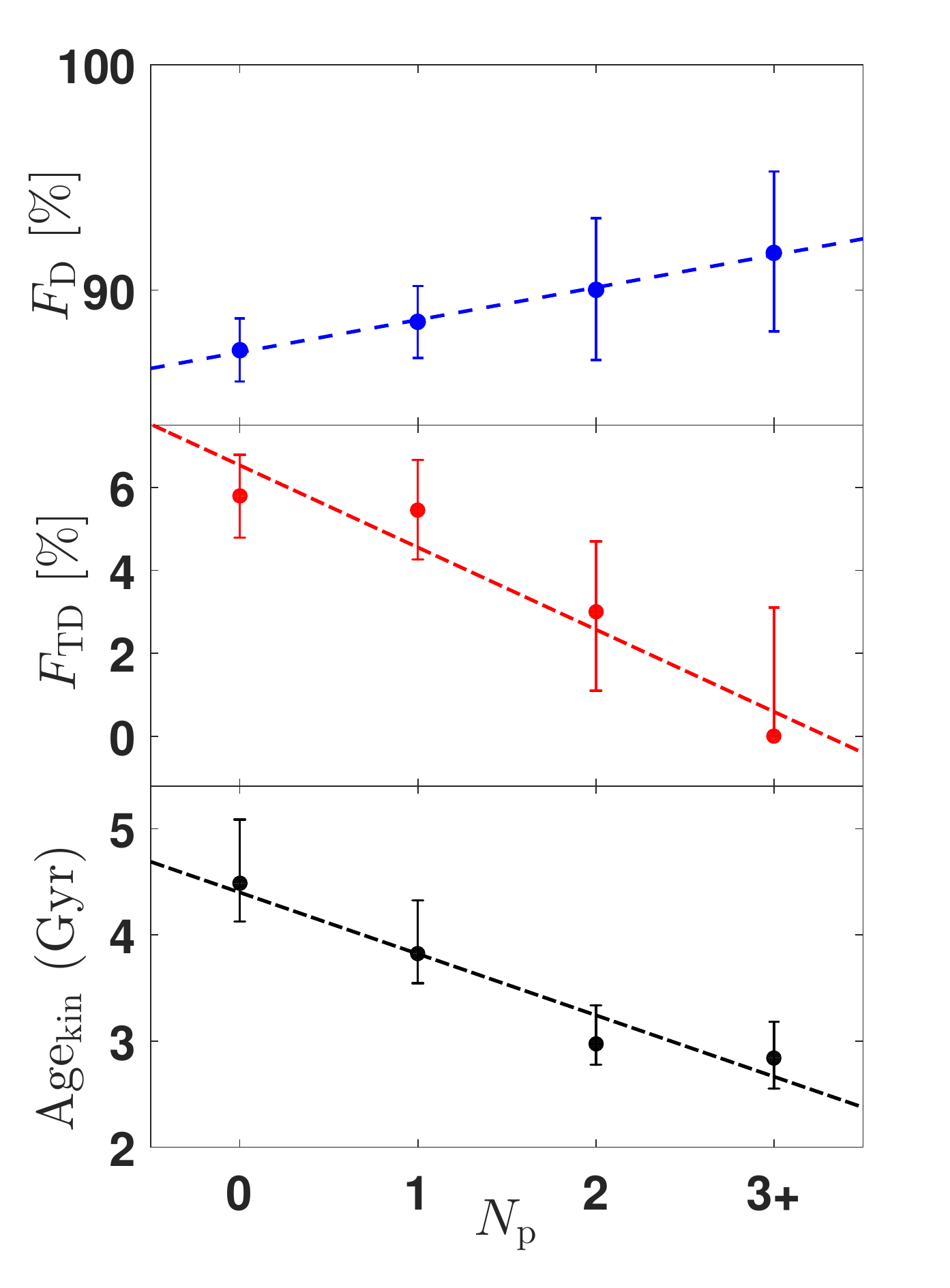}
\caption{\chendichang Fractions of thin disk stars ($F_D$, top panel) and thick disk stars ($F_{TD}$, middle panel) and kinematic age ($\rm Age_{kin}$, bottom panel) as functions of the the transiting planet multiplicity ($N_{\rm p}$) detected around a star in the LAMOST-Gaia-Kepler sample.
The dashed lines represent the best fits of linear model.
\label{figFDFTDtvsNp}}
\end{figure}


\subsubsection{Galactic components and kinematic age }
\label{sec.res.planethost.components}

Firstly, we compare the distributions of total velocity and the relative probabilities of the  thick-disk-to-thin-disk, $TD/D$ for Kepler {\chendichang planet candidate host} subsamples and neighbor non-Kepler {\chendichang planet candidate host} sample.
As can be seen in Figure \ref{figCDFvtotTDD}, the neighbor stars without Kepler planets (wide grey lines) have velocities and $TD/D$ that are on average larger but are
statistically indistinguishable (KS test $p-$values $>0.05$) compared to Kepler {\chendichang planet candidate host} stars (dashed red lines).
This is consistent with the result of \cite{2019MNRAS.489.2505M}, which found that Kepler {\chendichang planet candidate host}s (dominated by singles i.e., $N_{\rm p}=1$ since they are the majority) have similar Galactic velocity distribution with the Kepler field stars.
However, through further detailed exploration, we find that the total velocities and $TD/D$ generally decrease with the increase of the transiting planet multiplicity $N_{\rm p}$.
To quantify this trend, we did KS tests between the neighbor sample without Kepler planets and stars hosting 1, 2, $3+$ tranets systems. 
The resulting p-values {\chen (as printed in Figure 8)} decreases with increasing $N_{\rm p}$, demonstrating that the difference in the distributions of $V_{\rm tot}$ and $TD/D$ becomes greater and greater with the increase of $N_{\rm p}$.
Especially for stars hosting 2 and $3+$ tranets systems, the p-values are smaller than 0.05, demonstrating statistically smaller Galactic velocities and $TD/D$ (smaller fraction of thick disk).

Next, we compare the number fractions of different Galactic components. 
The number fractions of different Galactic components are calculated with the following formula:
\begin{equation}
F_i=\frac{N_i}{N_{D}+N_{TD}+N_{H}+N_{Herc}+N_{IB}},
\end{equation}
where $i$ represents the different Galactic components, i.e., thin disk (D), thick disk (TD), Halo (H), Hercules stream (Herc) and in between (IB).
To obtain the uncertainties, we assume that the observation numbers of different Galactic components obey the Poisson distribution.
Then we resample the observation numbers of different Galactic components from the given distribution and calculate the number fractions for 10,000 times.
The uncertainty (1--$\sigma$ interval) of each parameter is set as the range of $50 \pm 34.1$ percentiles of the 10,000 calculations.
The fractions of different components for the stellar sample without Kepler planets, Kepler {\chendichang planet candidate host} sample and the three subsamples are summarized in Table \ref{tab:starpropertywpnp}.
We find that with the increase of planet multiplicity ($N_{\rm p}$ from 0 to $3+$), the number fraction of the thin disk $F_{\rm D}$ increases, while the fraction of thick disk $F_{\rm TD}$ generally decrease, which are illustrated  in the upper and middle panels of Figure \ref{figFDFTDtvsNp}.

Then, we compare the kinematic ages of stars with and without Kepler planets. 
The kinematic ages and uncertainties are calculated from the velocity dispersions with Equation \ref{eqkineage} and \ref{eqkineageerr}.
As suggested in PAST \uppercase\expandafter{\romannumeral1}, there
is no clear trend between velocity dispersions and ages for stars belonging to Hercules stream and halo. Therefore we only consider stars with $Herc/D<0.5\&Herc/TD<0.5\&TD/H>1$ in calculating the kinematic ages. 
As shown in the bottom panel of Figure \ref{figFDFTDtvsNp}, the typical ages decrease from  {\chen $4.49^{+0.60}_{-0.36}$ Gyr, $3.82^{+0.50}_{-0.28}$ Gyr, $2.98^{+0.36}_{-0.20}$ Gyr to $2.84^{+0.34}_{-0.29}$ Gyr} when the number of planets $N_{\rm p}$ increases from 0 to $\ge 3$.
In 10,000 sets of resampled data, the kinematic age of neighbor non-Kepler {\chendichang planet candidate host} sample is bigger than that of $N_{\rm p}=1$, $N_{\rm p}=2$, and $N_{\rm p} \ge 3$ subsamples for {\chen 8,639, 9,981, and 9,995 times, corresponding to confidence level of 86.39\%, 99.81\% and 99.95\% respectively}.
We therefore conclude that the ages of 2 and $3+$ tranets systems are statistically smaller than stars without Kepler planets, while there is no significant difference for 1 tranet systems within the precision of our age estimations.
We summarized the median value and 1 $\sigma$ interval of physical and kinematic properties in Table \ref{tab:starpropertywpnp}.

Last but not least, we evaluate the statistical significance of the above trends as seen in Figure \ref{figFDFTDtvsNp}.
As can be seen in Figure \ref{figFDFTDtvsNp}, with the increase of $N_{\rm p}$, the number fraction of the thin disk $F_{\rm D}$ increases, while the fraction of thick disk $F_{\rm TD}$ and kinematic age $\rm Age_{kin}$ generally decrease.
To describe the property of the above trends mathematically,  we fit the relation between $F_D$, $F_{TD}$, $\rm Age_{kin}$ as a function of $N_{\rm p}$ with two models: constant model ($y=A_0$) and linear model ($y=A\times N_{\rm p}+b$).
For each model, we fit the relationships with the Levenberg-Marquardt algorithm (LMA). 
In order to compare the fitting of different models,  we calculate the Akaike information criterion \citep[AIC,][]{CAVANAUGH1997201} for each model. 
The AIC differences between the constant and linear model ($\Delta \rm AIC \equiv AIC_{con}-AIC_{lin}$) are {\chen 14.36, 7.69, and 9.15 for the trend of $F_D$, $F_{TD}$, and $\rm Age_{kin}$ respectively}.
Therefore a constant model can be confidently ruled out compared with the linear model with an AIC score difference $\Delta \rm AIC>6$.
To obtain the confidence level, we fit the relationships with LMA for the foregoing resampled data of $F_D$, $F_{TD}$ and $\rm Age_{kin}$.
Of the 10,000 sets  of resampling and fitting, the linear models are preferred with a smaller AIC score than the constant model for {\chen 9,525, 9,724 and 9,948 sets for the $F_D$, $F_{TD}$ and $\rm Age_{kin}$, corresponding to a confidence level of 95.25\%, 97.24\% and 99.48\%, respectively.

Furthermore, as shown in Figure \ref{fig_age}, kinematic age is overestimated as compared to asteroseismic age at the young end, i.e., $<2-3$ Gyr. If adopting such an age correction, the actual age of multi-planet systems (shown in bottom panel of Figure \ref{figFDFTDtvsNp}) should be smaller, which will make the trend even stronger. }

Based on above tests, we conclude that the trends between $F_D$, $F_{\rm TD}$, $\rm Age_{kin}$ and $N_{\rm p}$ as shown in Figure \ref{figFDFTDtvsNp} are statistically significant. The trends cannot be explained by the spatial bias and other effects related to stellar mass, radius, temperature,  and chemical abundances because, as can be seen in Table \ref{tab:starpropertywpnp} and Figure \ref{figCDFSpatial}, the Kepler {\chendichang planet candidate host} subsamples and the control sample all have similar distributions in spatial positions and physical properties. Therefore, we believe that these trends may be some footprints left in the evolution of planetary dynamics.

\section{summary}
\label{sec.summary}
The Kepler mission has detected over 4000 planets (candidates) by monitoring $\sim$ 200,000 stars. Accurately characterizing the properties of these stars is essential to study planet properties and their relations to stellar hosts and environments. Previous studies have investigated some of basic stellar properties, e,g., mass, radius \citep{2018ApJ...866...99B}, metallicity \citep{Dong.2013ApJ...778...53D}. 
However, the kinematic properties (e.g. Galactic component memberships and kinematic ages) of these stars are yet to be well characterized. 

In this paper, as the Paper \uppercase\expandafter{\romannumeral2} of the PAST series, we construct a LAMOST-Gaia-Kepler catalogue of {\chen 35,835} Kepler stars and {\chen 764} Kepler {\chendichang planet candidate host} stars with kinematic properties as well as other basic stellar properties (section \ref{sec.res.cat}, Table \ref{tab:starcatalog}) by combining data from Gaia DR2 and LAMOST DR4 (section \ref{sec.samp}, Table \ref{tab:planetnumberprocedure}).
By adopting {\chen the revised kinematic methods} in PAST \uppercase\expandafter{\romannumeral1}, we calculate the kinematics (i.e., Galactic position and velocity) and the Galactic component membership probabilities for the stars and then classify them into different Galactic components.
The fractions of {\chen stars belonging to} the thin disk, thick disk, Halo and Hercules stream are 87.1\% (31,218/{\chen 35,835}), 5.1\% (1,832/{\chen 35,835}), 0.16\% (59/{\chen 35,835}), and 1.5\% (545/{\chen 35,835}), respectively.
We also explore the kinematics and chemical abundances of different Galactic components.
{\chen As expected}, from the thin disk, Hercules stream, thick disk to the halo, the Galactic velocity is getting larger, the metallicity $\rm [Fe/H]$ decreases and $\rm [\alpha/Fe]$ increases (Table \ref{tab:starproperty} and Figure \ref{figVtotFealphaCDF}). 


Based on our LAMOST-Gaia-Kepler catalog, we derive the stellar kinematic ages with  typical uncertainties of 10-20\% by adopting the refined Age-velocity dispersion relation (AVR) of PAST \uppercase\expandafter{\romannumeral1} \citep{2021ApJ...909..115C}.
Comparing to ages derived from other methods (section \ref{sec.cat.analy.vetify}), we find kinematic age generally matches with asteroseismic, gyrochronology, and isochrone ages  (Figure \ref{fig_age}). 
We have carried out some analyses to explore the connection between stellar activity and kinematic age (section \ref{sec.cat.stellaractivity}). 
As expected, {\chen as stars age}, they spin down and become less active in terms of both the magnetic S index (Figure \ref{fig_rot_rhk}) and the photometric variability (Figure \ref{fig_cdpp}), which, in turn, verify the kinematic ages derived in this work.

We have also explored the differences in kinematic properties (e.g., Galactic velocities, thin/thick disk memberships and kinematic ages)  between the Kepler {\chendichang planet candidate host}s and stars without Kepler planets (section \ref{sec.res.planetvsnonplanethost}).
For the aspect of spatial position, the stars without Kepler planets have a wider distribution in the distance to the Sun, Galactic radius ($R$), azimuth angle ($\theta$) and height ($Z$) than that of Kepler {\chendichang planet candidate host} stars (Figure \ref{figCDFSpatial}), which could be an observational selection bias as planet systems closer to us are easier to be detected \citep[e.g.][]{2019MNRAS.489.2505M}.
Thus for a fair comparison between the kinematic properties of stars with and without Kepler planets, we
construct a control sample by selecting the nearest neighbors of {\chendichang planet candidate host}s in the spacial distribution (Figure \ref{figCDFSpatial}). 
Comparing to stars in the control sample, we find that {\chendichang planet candidate host}s, especially those with large planet multiplicity ($N_{\rm p}\ge3$), differ significantly in  distributions of the Galactic velocity $V_{\rm tot}$ and the relative probability between thick and thin disks $TD/D$ (Figure \ref{figCDFvtotTDD}).
In particular, we find a trend that the fraction of thin (thick) disk stars increases (decreases) with transiting planet multiplicity ($N_{\rm p}$) and the kinematic age decreases with $N_{\rm p}$  (Figure \ref{figFDFTDtvsNp} and Table \ref{tab:starpropertywpnp}).
This provides insights into the formation and evolution of planetary systems with the Galactic components and stellar age.
One possible explanation for the trend is that the long term dynamical evolution can pump up orbital eccentricity/inclination of planets \citep[e.g.][]{2007ApJ...666..423Z} or even cause planet merge/ejection \citep[e.g.][]{2015ApJ...807...44P}, which reduces the observed transiting planet multiplicity i.e., $N_{\rm p}$.
Specifically, in a subsequent paper of the PAST project (Yang et al. in prep), we will study whether/how planetary occurrence and architecture (e.g. inclination) change with the ages and Galactic environments based on the LAMOST-Gaia-Kepler catalog of this work.



The LAMOST-Gaia-Kepler catalog provides the kinematics, Galactic component-memberships, $\rm [Fe/H]$, $\rm [\alpha/Fe]$ and ages information for thousands of planets (candidates) down to about Earth radius and tens of thousands of well-characterized field stars with no bias toward Kepler {\chendichang planet candidate host}s, which will be useful for more future studies of exoplanets at different positions/components of the Galaxy with different ages.
The answers of these questions will deepen our understanding of planet formation and evolution.

\section*{Acknowledgements}
This work has included data from Guoshoujing Telescope
(the Large Sky AreaMulti-Object Fiber Spectroscopic Telescope LAMOST), which is a National Major Scientific Project built by the Chinese Academy of Sciences. Funding for the project has been provided by the National Development and Reform Commission.
This work presents results from the European Space Agency (ESA) space mission Gaia. Gaia data are being processed by the Gaia Data Processing and Analysis Consortium (DPAC). Funding for the DPAC is provided by national institutions, in particular the institutions participating in the Gaia MultiLateral Agreement (MLA). The Gaia mission website is https://www.cosmos.esa.int/Gaia. The Gaia archive website is https://archives.esac.esa.int/Gaia.
We acknowledge the NASA Exoplanet archive, which is operated by the California Institute of Technology, under contract with the National Aeronautics and Space Administration under the Exoplanet Exploration Program.

This work is supported by the National Key R\&D Program of China (No. 2019YFA0405100) and the National Natural Science Foundation of China (NSFC; grant No. 11933001, 11973028,  11803012, 11673011, 12003027). J.-W.X. also acknowledges the support from the National Youth Talent Support Program and the Distinguish Youth Foundation of Jiangsu Scientific Committee (BK20190005).
C.L. thanks National Key R\&D Program of China No. 2019YFA0405500 and the National Natural Science Foundation of China (NSFC) with grant No. 11835057.
H.F.W. is supported by the LAMOST Fellow project, National Key Basic R\&D Program of China via 2019YFA0405500 and funded by China Postdoctoral Science Foundation via grant 2019M653504 and 2020T130563, Yunnan province postdoctoral Directed culture Foundation, and the Cultivation Project for LAMOST Scientific Payoff and Research Achievement of CAMS-CAS.
M. Xiang \& Y. Huang acknowledge the National Natural Science Foundation of China (grant No. 11703035).

\bibliography{library.bib}

\end{document}